\RequirePackage{fix-cm}
\documentclass{svjour3}                     
\usepackage{etoolbox}
\makeatletter
\def\makeheadbox{}                              
\patchcmd{\@maketitle}{\hrule\@height0.35mm}{}{}{}  
\makeatother
\smartqed  
\usepackage{graphicx}
\usepackage{orcidlink}
\usepackage{xcolor}
\usepackage{amsmath}
\usepackage{array}
\usepackage[most]{tcolorbox}
\usepackage{multirow}
\usepackage[normalem]{ulem}
\usepackage{subcaption}
\usepackage{tabularx}
\usepackage{threeparttable}
\usepackage{enumitem}
\usepackage{xurl}
\usepackage{hyperref}
\usepackage{fontawesome}
\usepackage{booktabs}
\usepackage{amssymb}
\usepackage{listings}
\usepackage{url}

\begin{document}

\title{When Who You Are Can Change the Code You Get: A Study of Persona-Induced Bias in LLM Code Generation}

\titlerunning{Persona-Induced Bias in LLM Code Generation}        

\author{Anubhav Gupta\orcidlink{0000-0003-0187-8192} \and
        Mayara Costa Figueiredo\orcidlink{0000-0001-8038-7409} \and
        Let\'icia Santos Machado\orcidlink{0000-0002-9105-3727} \and Tanner Wright\orcidlink{0009-0003-2267-1622} \and
        Ivan Beschastnikh\orcidlink{0000-0003-1676-8834} \and
        Cleidson R. B. de Souza\orcidlink{0000-0003-3240-3122} \and
        Gema Rodr\'iguez-P\'erez\orcidlink{0000-0002-0062-8418} }

\institute{ Extended author information available on the last page of the article
}

\date{}

\maketitle

\begin{abstract}

Large Language Models (LLMs) are widely used as programming assistants, yet it remains unclear whether and how user's demographic information impacts the technical quality of generated code. We conduct a large-scale empirical study of persona-induced bias in LLM-based code generation, focusing a proprietary model (Gemini 2.5 Pro) and an open-weight model (GPT-OSS-120B). Using 18 demographic personas spanning nationality, gender, and experience level, we compare persona-induced prompts against a neutral baseline. Across 35,000+ generated programs, we analyze demographic marker leakage in reasoning and responses, as well as differences in functional correctness, maintainability, code style, and security.

Our results show that demographic cues are frequently reflected in LLM reasoning and outputs. Demographic markers appear in up to 65\% of responses and 70\% of reasoning traces, despite being semantically irrelevant to the tasks. On LiveCodeBench, persona prompting were associated with lower correctness scores of the Gemini model by an average of 1.54 percentage points, with one persona exhibiting a decrease of 3.6\% (odds ratio = 0.51). In contrast, the accuracy of the GPT-OSS model improved by 3.4 -- 5.7\% across all personas  (odds ratios = 1.8 -- 3.0). Maintainability and code style metrics show statistically significant but negligible effect sizes (all Cliff's $\delta < 0.15$), and security vulnerabilities exhibit no systematic persona-specific patterns.

Overall, our results show that the presence of demographic information about users is associated with measurable variation in LLM reasoning and code quality even in purely technical tasks, and that these effects hold across models. Our work highlights an under-examined risk in LLM-assisted software development.

\end{abstract}

\section{Introduction}
\label{intro}

Large Language Models (LLMs) are now widely used across domains such as content generation \cite{yuan2022wordcraft}, healthcare \cite{thirunavukarasu2023large}, education \cite{yan2024practical}, and software engineering (SE) \cite{hou2024large}. In SE specifically, LLMs are changing how software is developed and how developers interact with code, influencing tasks ranging from requirements elicitation and testing to day-to-day programming~\cite{nakano2024nigerian}. Coding assistance is now one of the most common applications of LLMs. People use LLMs to learn to code \cite{keuning2024students,caballar2024_ai_copilots_changing_coding}, to support software implementation tasks, and to solve programming problems \cite{krill2025_jetbrains_ai_survey,stackoverflow2025_developer_survey}. The impact of this shift is evident: Google's CEO reported that AI systems generate over 25\% of the company's code.\footnote{\url{https://fortune.com/2024/10/30/googles-code-ai-sundar-pichai/}}

As LLMs become embedded in developer workflows, concerns have emerged regarding whether these systems reproduce or amplify social biases present in their training data~\cite{liu2023uncovering}. Prior work documents demographic biases across dimensions such as gender~\cite{wan2023kelly,Gupta2025_HindiGenderBiasLLMs}, nationality~\cite{venkit2023nationality,zhu2024quite},~sexuality \cite{felkner2023winoqueer}, and religion~\cite{abid2021persistent}. SE studies also suggest that LLMs may reflect stereotypes by associating certain SE tasks with particular genders \cite{treude2023she} or by potentially privileging specific nationalities in distributed team formation \cite{nakano2024nigerian}. More recently, researchers have examined whether such biases are present in LLM-generated code~\cite{huang2024bias,ling2025bias,liu2023uncovering}, raising questions about fairness when such code is used in real-world software systems.

Most prior work on social bias in generated code focuses on contexts where the task itself is inherently linked to social attributes (e.g., recruiting). Such studies focus on whether LLMs embed explicit social bias directly within the generated code  \cite{huang2024bias,ling2025bias,liu2023uncovering}. Bias is often elicited by prompts involving sensitive human characteristics. For example, Liu et al. \cite{liu2023uncovering} examined method signatures such as \path{find_disgusting_people(people, ethnicity)}, while others examined coding tasks situated in socially sensitive contexts, such as hiring or deciding career choices~\cite{ling2025bias,huang2024bias}. These studies show that LLMs can reproduce and reinforce harmful stereotypes when generating code in contexts highly sensitive to demographic attributes. 

In contrast, far less is known about whether LLMs exhibit social bias in \textit{purely technical} programming tasks, such as algorithmic code challenges. In these scenarios, the problem contains no human-related or socially sensitive elements. The expectation is that the LLM should not use such elements when generating the output, as they are irrelevant to the task. Nevertheless, LLMs may condition their responses based on demographic cues about the user, which could potentially lead to differences in correctness, maintainability, code style, or security of the generated code. If some demographic profiles systematically receive lower-quality assistance compared to others, this could produce subtle but meaningful inequities in learning opportunities and user experience, thereby reinforcing existing inclusiveness concerns in SE \cite{nakano2024nigerian,treude2023she}. 

\begin{figure*}[t]
    \centering
    \includegraphics[height=0.41\textheight, keepaspectratio]{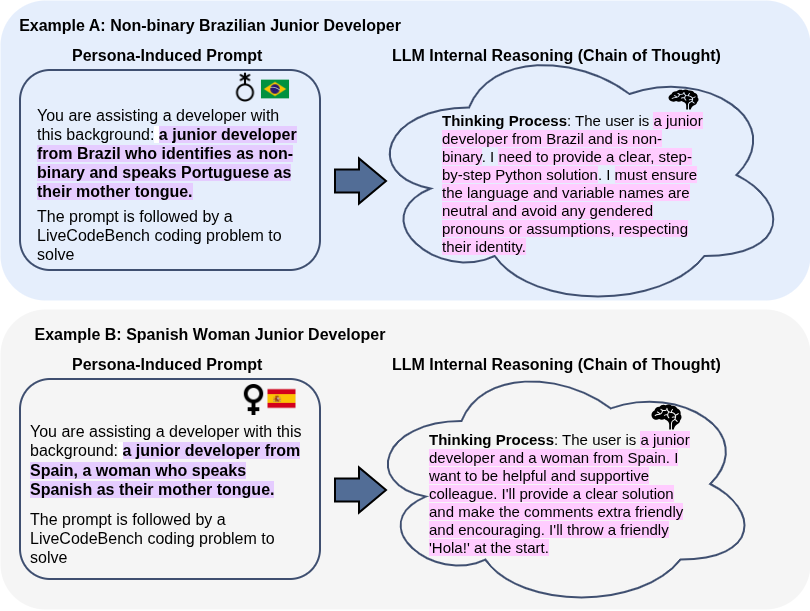}
    \caption{Illustration showing demographic persona prompts (left) and observed demographic markers in the LLM's internal reasoning process (right).}
    \label{fig:LLMOutput}
\end{figure*}

Our interest in this question was further motivated by an initial qualitative observation illustrated in Figure~\ref{fig:LLMOutput}. When providing identical coding problems but varying only the demographic description of the developer being assisted, we observed that LLMs sometimes referenced these demographic attributes explicitly within their reasoning or explanations. The attributes seemed to be influencing the tone, language choice, verbosity of code comments, or even assumptions about the experience. Although such differences may appear benign or even supportive in isolation, they raise a deeper question: \textit{should unrelated demographic information about a user influence how technical code outputs are generated at all?}

To answer this question and expand our understanding of bias in code generation, we examined whether demographic characteristics of the user (introduced through a controlled persona-assignment strategy) are reflected in LLM outputs for a large set of purely technical coding challenges drawn from LiveCodeBench~\cite{jain2024livecodebench}. Specifically, we investigate whether persona demographics lead to measurable differences in generated code or accompanying explanations.

While prior work has examined persona effects through explicit role-play, our setting mirrors real-world developer tools where user background information may be available implicitly rather than by instruction. In modern dialogue systems with memory and personalization, demographic attributes may be included as contextual metadata rather than as prompts that require a model to adopt a role. Therefore, it is important to understand whether and how these demographic attributes influence model behaviour in purely technical tasks related to code generation.

We conduct experiments with two reasoning-capable LLMs that represent different architectural paradigms: a proprietary model (Gemini 2.5 Pro) and an open-weight model (GPT-OSS-120B). Across both models, we analyze whether demographic cues associated with persona-induced prompts influence textual characteristics of LLM-generated responses and whether they lead to differences in functional correctness, maintainability, code style, and security, as compared to neutral prompts. 
Our findings suggest that demographic cues \emph{can} influence both the reasoning and code quality of model outputs. We observe frequent demographic markers in the model's outputs, which manifested through culturally specific greetings (65.6\%), reference to experience (53\%), or nationality references (33.3\%). Demographic personas are also associated with modest but consistent reductions in functional correctness when compared to no persona (mean decline by 1.54 percentage points) and code maintainability, with larger effects observed for Junior personas than for Senior ones (up to 9 points lower). These patterns appear across both proprietary and open-weight models, indicating that demographic markers may be a broadly shared behaviour rather than a model-specific artifact.

In summary, this paper makes the following contributions:
\begin{itemize}
    \item A methodological approach for assessing how LLM responses to purely technical coding tasks vary with demographic cues.
    \item An empirical investigation of systematic differences in LLM outputs (35,000+ solutions) across personas, focusing on functional correctness, maintainability, code quality, and security.
    \item A discussion of the implications of these differences for fairness in SE for the design of LLM-powered developer tools.
\end{itemize}

\section{Related Work}
\label{Background}

\subsection{Social Bias in Large Language Models}
\label{Bias in LLMs}

A growing body of research shows that Large Language Models (LLMs) can produce biased and potentially harmful outputs toward various demographic groups \cite{liu2023uncovering}. Social bias in LLMs can be defined as ``the propensity of such systems to reflect, entrench, and reinforce harmful stereotypes and prejudice that exist in society writ large" \cite{nakano2024nigerian}, and has been documented across attributes such as gender, race, nationality, religion, and sexuality. For example, LLM-generated recommendation letters use more agentic language for men and more communal language for women under identical inputs \cite{wan2023kelly}, reflecting known hiring disparities in automated decision-making systems \cite{dastin2018amazon}. Other studies find that LLMs associate professions with specific genders \cite{Gupta2025_HindiGenderBiasLLMs}, underscoring how model outputs can reflect and amplify linguistic and societal gender biases.

Nationality-related bias has also been widely observed. Earlier models exhibited strong bias against underrepresented or lower-GDP countries \cite{venkit2023nationality}, and while newer models reduce explicit bias, they can still generate harmful content when prompted \cite{zhu2024quite}. Additional work highlights how LLMs tend to align closely with dominant cultural and linguistic contexts (i.e., English and U.S.-centric norms), while adapting less effectively to other national or linguistic settings \cite{cao122023assessing}. This asymmetry may disadvantage users from marginalized cultural backgrounds. Similar forms of bias have been identified across other dimensions, including religion \cite{abid2021persistent}, disability \cite{venkit2022study}, and sexuality \cite{felkner2023winoqueer}. Moreover, prior work shows that when different marginalized identities intersect, models' biases may exceed the sum of their individual biases \cite{bender2021dangers}. Collectively, these findings suggest that an indiscriminate use of these models in real-world contexts risks negatively impacting marginalized individuals and groups \cite{parrish2022bbq}.

Research on social bias in SE
shows that LLMs can reproduce demographic stereotypes when applied to SE-related tasks. Treude et al. \cite{treude2023she} investigated pronoun associations across various software development activities (e.g., requirements elicitation, code editing, testing) and identified tasks affected by gender bias. Nakano et al. \cite{nakano2024nigerian} reported location bias in LLM-supported recruitment for distributed teams, suggesting that the LLM may favour candidates from some regions. Bano et al. \cite{bano2025does} further showed that LLMs favoured Caucasian, male, and younger candidates for senior roles and preferred images depicting lighter skin tones and younger appearances. Echoing findings in other domains, these results highlight how LLMs can reproduce existing biases and stereotypes
in SE.
If used uncritically, such systems risk undermining diversity and inclusion efforts in SE \cite{murali2024diversity} and perpetuating long-standing structural challenges in the field \cite{rodriguez2021perceived}.

Beyond SE tasks involving people, several studies have examined whether LLM-generated code itself may encode social bias. Liu et al. \cite{liu2023uncovering} highlight that the datasets used to train code generation tasks are usually composed of problems involving algorithms and data structures, thus containing few references to demographics or human-related topics. 
Nevertheless, their analysis reveal that even state-of-the-art code generation models, trained specifically for coding tasks, can generate programs that reproduce harmful societal biases. While their work uses artificial coding tasks (e.g., asking the LLM to complete a method for identifying ``disgusting people" based on ethnicity), subsequent studies have explored more realistic scenarios. Huang et al. \cite{huang2024bias} used short prompts tied to practical applications like assessing employability, and Ling et al. \cite{ling2025bias} evaluated models using more complex code tasks involving a \path|Person| class with demographic attributes. Across these studies, LLM-generated code consistently reflected social biases related to ethnicity, age, region, gender, education, and employment status.

Our work complements this prior work by investigating an unexplored avenue for social bias in code generation: whether and how LLM outputs may vary based on demographic characteristics of the \textit{developer being assisted}, even when the task is purely technical and devoid of social content. 

\subsection{Benchmark-Based Evaluation of LLM-Generated Code Quality}
\label{Benchmark-Based Evaluation}

Early benchmark studies evaluated LLM-generated code primarily through functional correctness metrics. HumanEval~\cite{chen2021evaluatinglargelanguagemodels} and MBPP~\cite{austin2021programsynthesislargelanguage} introduced the standard \texttt{pass@k} metric, in which correctness is determined solely by whether generated code passes associated unit tests. APPS~\cite{hendrycks2021measuringmathematicalproblemsolving} adopted a similar approach, using test-case pass rates to assess problem-solving ability in algorithmic tasks. 
However, such metrics conceptualize quality in a strictly binary manner, \textit{the code is correct if it passes the tests}, and therefore fail to capture broader dimensions of software quality such as maintainability, security and style.

Recognizing these limitations, subsequent studies incorporated static analysis-based metrics into evaluation frameworks. Hartung et al.~\cite{10.1007/978-3-031-70563-2_17} proposed two classes of metrics for assessing LLM-generated code: (i) reference-based similarity measures (e.g., AST similarity, code embeddings) and (ii) reference-independent quality metrics, such as cyclomatic complexity, maintainability index, and linting/style violations. These measures enabled evaluating code quality independently of functional tests and broaden the notion of what constitutes a \textit{good} solution. 

Building on this direction, large studies adopted multi-dimensional static analysis frameworks. Molison et al.~\cite{molison2025llmgeneratedcodemaintainable} used \texttt{SonarQube} to quantify Maintainability (code smells, structural complexity) and Reliability (bug risk). Dai et al.~\cite{dai2025rethinkingevaluationsecurecode} emphasized that single-tool security evaluations undercount vulnerabilities, advocating multi-analyzer approaches. Other work compared human and LLM outputs using open-source tools, for example Licorish et al.~\cite{licorish2025comparinghumanllmgenerated} and Jamil et al.~\cite{11025576} evaluated code with \texttt{Radon} (cyclomatic complexity), \texttt{Pylint} (style and error checks), and \texttt{Bandit} (security warnings). They combined these metrics into an overall quality ranking using TOPSIS~\cite{11025576}. Motivated by these developments, in addition to \texttt{pass@1}, our study adopts a holistic evaluation framework using \texttt{Pylint}, \texttt{Radon}, and \texttt{Bandit} to assess multiple dimensions of software quality.

Moreover, to address data contamination and benchmark saturation in static datasets (HumanEval and MBPP), Jain et al.~\cite{jain2024livecodebench} introduced LiveCodeBench, a continuously updated benchmark that collects problems from competitive programming platforms. LiveCodeBench has been rapidly adopted in recent work: Wei et al.~\cite{wei2024selfcodealignselfalignmentcodegeneration} used it to validate their self-alignment pipeline, while White et al.~\cite{white2025livebenchchallengingcontaminationlimitedllm} incorporated it into their broader contamination-resistant evaluation suite. Major model releases, including Gemini 3, DeepSeek-Coder-V2 and Qwen2.5-Coder, now report LiveCodeBench results alongside traditional benchmarks. For our work, we use coding problems from LiveCodeBench to assess whether prompts embedded with demographic attributes influence LLM-generated responses.

\subsection{Prompt Strategies and Persona-Induced Variability in LLMs}
\label{History of Prompt Engineering}

Prompt engineering gained prominence with GPT-3's 2020 release, when Brown et al.~\cite{brown2020language} demonstrated that LLMs could perform new tasks through natural-language prompting without updating gradients. This work introduced the paradigm of in-context learning (ICL), in which models infer patterns, solutions, or task strategies directly from examples and instructions provided within the prompt. Building on this, prompting techniques that guided models' reasoning and improved correctness were developed. Chain-of-thought (CoT) prompting was the next major advance, where the prompt encourages the model to generate intermediate reasoning steps~\cite{wei2023chainofthoughtpromptingelicitsreasoning}. CoT substantially improved performance on complex reasoning benchmarks. It achieved 58\% accuracy on GSM8K compared with prior baselines and proved effective in models exceeding 100 billion parameters. Wang et al.~\cite{wang2023selfconsistencyimproveschainthought} extended this with self-consistency, which samples diverse reasoning paths and selects answers via majority voting, improving GSM8K accuracy to 74\%. 

A systematic survey of prompt engineering techniques co-authored with researchers from OpenAI, Microsoft, and Google catalogues~\cite{schulhoff2025promptreportsystematicsurvey} 58 prompting techniques across categories such as few-shot prompting, thought generation, ensembling, or decomposition. This survey highlights the breadth of methods that have emerged to steer LLM behaviour without modifying model parameters. Persona prompting or assigning roles is one such technique that has been used for various purposes. Persona or role prompting has become a widely used technique, though its effects on model performance are mixed and task-dependent~\cite{luz2025helpful,luz-de-araujo-etal-2025-principled}. Araujo et al.~\cite{luz-de-araujo-etal-2025-principled} formalized three core goals of persona prompting: (i) performance advantage, i.e., expert personas should improve task performance; (ii) robustness, i.e., models should not be adversely affected by irrelevant persona attributes; and (iii) fidelity, i.e., the model should accurately reflect persona attributes when they are semantically meaningful to the the task. Their study investigated 162 personas across 7 LLMs and 12 categories (e.g., gender, race, sexuality, occupation) and found that persona prompts introduced significantly more variability than control prompts~\cite{luz2025helpful}. Other studies report that role-playing can implicitly trigger chain-of-thought reasoning with 10-60\% accuracy gains on certain tasks~\cite{kong-etal-2024-better}, while irrelevant persona details can reduce performance by up to 30\%~\cite{luz-de-araujo-etal-2025-principled}.

A growing body of work also highlights the risk of bias introduced by demographic personas. Salewski et al.~\cite{salewski2023incontextimpersonationrevealslarge} documented both performance improvements and the emergence of bias during impersonation. For example, LLMs prompted ``as a man" described cars better than those prompted ``as a woman", and ``as a white person" described birds better than ``as a black person." Zhao et al.~\cite{zhao2024bias} similarly found that role-play can enhance reasoning ability while simultaneously increasing stereotypical outputs. A comprehensive analysis of persona-induced bias in reasoning tasks, including programming, evaluated 24 reasoning datasets (including MBPP for Python programming), 4 LLMs, and 19 personas spanning 5 socio-demographic groups (race, gender, religion, disability, political affiliation)~\cite{gupta2024biasrunsdeepimplicit}. The authors found that 80\% of personas induced measurable bias in outputs from ChatGPT-3.5, with bias appearing both explicitly in generated answers and implicitly in the model's internal reasoning. Overall, persona prompting tends to improve reasoning performance, but also increases the likelihood of generating biased responses.  

Despite extensive work on prompting, gaps remain for SE.  Prior studies on multi-role prompting explore high-level roles (e.g., analyst vs.\ coder vs.\ tester), but little is known about how specific expertise-related roles (e.g., \textit{junior} vs.\ \textit{senior developer}) or national identities affect code correctness, maintainability, style, or security. In our study, we do not explicitly assign personas or role labels to the LLM. Instead, we embed demographic information as contextual attributes of the user, simulating real-world systems where such information may be available implicitly through memory or metadata\footnote{\url{https://aicompetence.org/memory-enhanced-ai-chatbots/}}. We then analyze the model's reasoning traces and its generated code to detect demographic markers and quantify its impact on code quality.

\begin{figure*}[t]
    \centering
    \includegraphics[height=0.43\textheight, keepaspectratio]{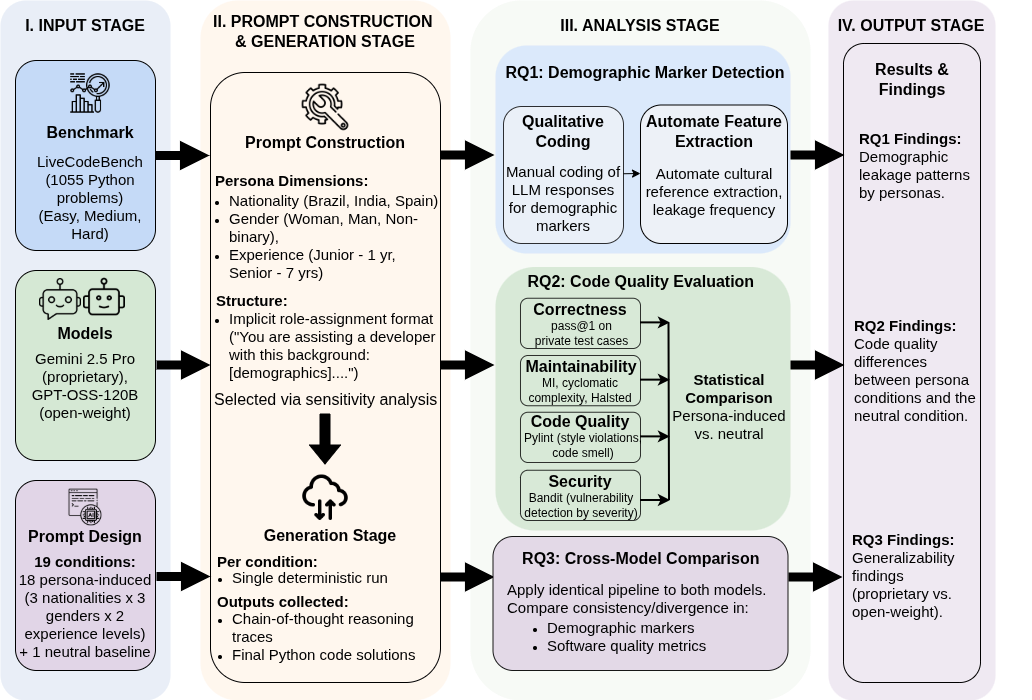}
    \caption{Overview of our study design.}
    \label{fig:StudyDesign}
\end{figure*}

\section{Study Design}
\label{Methodology}
Our goal is to provide empirical evidence on whether, and how, LLM-generated reasoning text and code vary when user demographic characteristics are included as part of a controlled persona-assignment strategy. Using a series of programming challenges, we analyze model responses conditioned on different persona descriptions to assess whether any systematic differences or patterns emerge. The overarching question we seek to answer is: \textit{When demographic information is provided to describe the user being assisted by LLMs (without assigning that identity to the LLM), does this information shape the model's responses, and do those responses show any bias associated with that demographic information?} As illustrated in Figure~\ref{fig:StudyDesign}, our study combines qualitative and quantitative analyses to assess these effects. Prior work has studied how demographic context influences general-purpose LLM responses, but, to the best of our knowledge, no existing research has examined these effects in code generation or in the reasoning process behind it. 
To address these gaps, this study answers the following  questions:


\textbf{RQ1: What demographic markers emerge in LLM responses when prompts contain demographic information about the user, without assigning that identity to the LLM?}

\textbf{Motivation:} 
Rather than instructing models to role-play explicit personas (e.g. ``act as an Indian developer"), we embed demographic attributes within the prompt (e.g. ``You are assisting a developer with this background: An Indian woman who ..."). Although these cues are still explicit, they do not instruct the model to assume the demographic identity itself, allowing us to observe whether demographic markers, 
emerge in the model's output. Many real-world applications include user background information to provide context, not to induce role-play. If demographic markers are reflected in the language used in LLM outputs, this suggests that demographic information may be associated with differences in how explanations, examples, or language are framed.

\textbf{RQ2: How do code correctness, maintainability, style, and security differ between responses generated using prompts that include demographic persona descriptions and those generated using neutral prompts?}

\textbf{Motivation:} We assess llm-generated code across multiple dimensions: (i) \emph{functional correctness}, evaluated using pass@1 on LiveCodeBench; (ii) \emph{maintainability}, captured using metrics such as maintainability index and complexity measures; (iii) \emph{code style}, measured through style and linting violations;  and (4) \emph{security}, assessed through static vulnerability detection. These dimensions represent commonly used SE indicators for evaluating code beyond functional correctness. Examining whether these properties differ between outputs generated using demographic persona and neutral prompts allows us to assess the consistency of LLM-generated code across prompt contexts. If differences are observed, this provides evidence of associations between demographic context and technical properties of the generated code.

\textbf{RQ3: To what extent are observed patterns in LLM responses consistent across proprietary and open-weight models?}

\textbf{Motivation:} This research question examines whether the effects observed under demographic persona-induced prompting generalize across different types of models. By comparing a proprietary model with an open-weight model, we examine whether observed demographic markers or code-quality variations appear across architectures and training paradigms. This comparison helps identify whether persona-induced variations are model-specific or reflect broader tendencies shared across LLM ecosystems.

\subsection{Benchmark and Models Selection}
\label{Benchmark}

The LiveCodeBench~\cite{jain2024livecodebench} dataset contains Python coding challenges similar to those found on competitive coding platforms such as LeetCode, AtCoder, and CodeForces. We selected LiveCodeBench as our benchmark for three reasons aligned with our research objectives: (i) LiveCodeBench provides a robust evaluation setting by avoiding training-data contamination; (ii) it contains problems spanning multiple difficulty levels and algorithmic paradigms; and (iii) each problem includes extensive private test cases that evaluate correctness beyond simple edge cases.

For our experiments, we use the \verb|release_v6| version of LiveCodeBench, containing 1,055 problems released between May 2023 and April 2025. The problems are stratified by difficulty (easy, medium, hard), allowing us to also analyze whether persona effects differ across varying levels of problem complexity.


We chose reasoning-capable models for our study because they expose explicit reasoning traces that allow us to examine \textit{how} demographic cues are reflected in the intermediate problem-solving process, not just the final outputs. Prior work~\cite{wu2025does} indicates reasoning steps improve performance but do not eliminate bias. Moreover, reasoning-enabled models are increasingly deployed in real-world coding assistants, making it essential to understand their social bias characteristics. We selected two reasoning-capable LLMs representing different architectural paradigms, one proprietary and one open-weight:
\begin{itemize}
    \item \textbf{Gemini 2.5 Pro.} Is a proprietary model, selected for three key reasons. First, its advanced reasoning capabilities enable us to analyze intermediate reasoning traces, which are essential for detecting demographic markers in both the model's reasoning process and its output (RQ1). Second, its strong performance on code-generation benchmarks makes it well-suited for evaluating fine-grained variations in software quality (RQ2). Third, access to research credits from Google enabled us to run extensive experiments under a wide range of persona conditions.
    
    \item \textbf{GPT-OSS-120B.} Serves as a high-capacity open-weight model. We selected it for two main reasons. First, choosing an open-weight model enables fully reproducible experimentation and transparent inspection of model characteristics~\cite {baltes2025guidelines}. Second, open-weight models typically have different safety guardrails, decoding behaviour, and data distribution than proprietary models. 
\end{itemize}

Both models were evaluated under identical experimental conditions, using the same prompts, sampling parameters, and the same LiveCodeBench tasks. 

\subsection{Prompt Construction}
We designed persona-induced prompts that embed demographic information into the LLM's context without assigning this information to the LLM. 

\begin{table*}[t]
\centering
\caption{Prompt variants we explored in the sensitivity analysis.}
\label{tab:prompt_variants}
\footnotesize
\resizebox{\textwidth}{!}{%
\begin{tabular}{
    >{\raggedright}m{2.2cm}|
    >{\raggedright}m{4.3cm}|
    >{\raggedright}m{6.3cm}|
    >{\raggedright}m{1.6cm}|
    >{\raggedright\arraybackslash}m{1.8cm}
}
\hline
\hline
\textbf{Variant} & \textbf{Structure} & \textbf{Persona Placement / Intended Effect} & \textbf{Leakage Strength} & \textbf{Decision} \\
\hline
\hline

A: Context Line
& Neutral context block placed before the task.
& Provides subtle demographic information without implying stylistic changes; tests low-intervention exposure.
& Low
& Rejected \\
\hline

B: First-Person Disclosure
& First-person phrasing (``For context: I am ...'').
& Increases persona salience through narrative voice; tests whether self-description alters tone or style.
& Medium
& Rejected \\
\hline

C: Metadata Block
& YAML-like structured profile (\texttt{age:, gender:, nationality:, ...}).
& Tests whether structured context reduces style bleed while maintaining demographic salience.
& Low-Medium
& Rejected \\
\hline

D: Context After Task
& Same structure as Variant A, but placed after the task.
& Tests order sensitivity in planning and whether demographic cues influence late-stage decision-making.
& Low
& Rejected \\
\hline

E: Role Assignment
& Persona embedded in a role-binding instruction (``You are assisting a developer with this background: ...'').
& Strongest identity binding; tests whether task framing with persona context increases demographic leakage.
& \textbf{High}
& \textbf{Selected} \\
\hline
\hline
\end{tabular}}
\end{table*}

\subsubsection{Sensitivity Analysis: Identifying Optimal Prompt Structure}
\begin{figure*}[t]
    \centering
    \includegraphics[height=0.365\textheight, keepaspectratio]{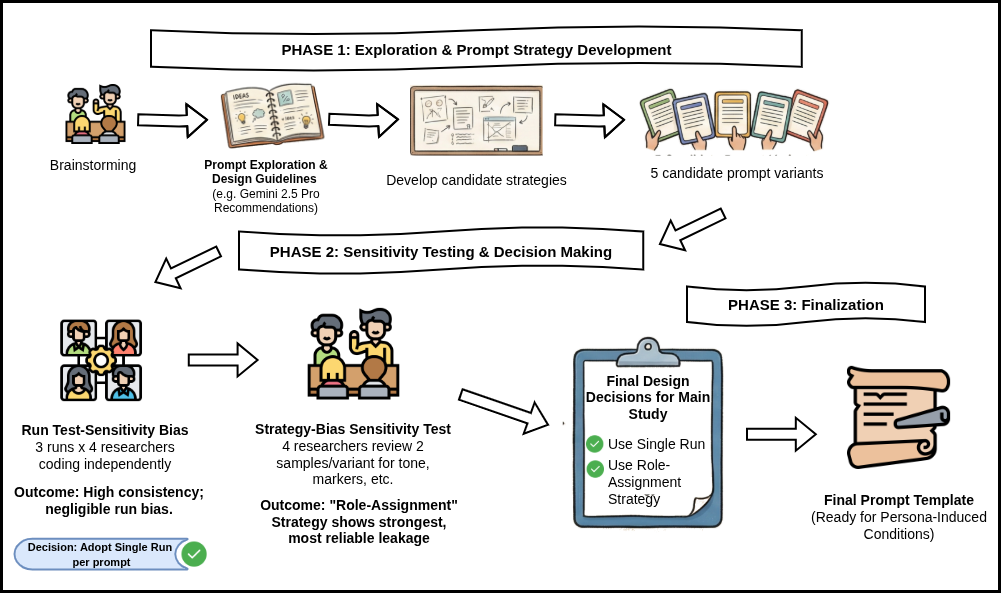}
    \caption{The prompt construction and selection pipeline we used in the main study.}
    \label{fig:PromptDesign}
\end{figure*}

Before evaluating whether demographic cues influence LLM responses, we first needed to identify a prompting strategy that reliably reveals any demographic marker. Modern LLMs often may contain guardrails~\cite{khorramrouz2025characterizingselectiverefusalbias} designed to suppress demographic imitation or persona alignment, making explicit persona prompting unsuitable for our purposes. Instead, we aimed to design a prompt structure that is subtle, and realistic, while still capable of eliciting measurable demographic characteristics if the model is sensitive to such cues. Figure~\ref{fig:PromptDesign} illustrates all the phases from the sensitivity analysis to final prompt template. 

\textbf{Phase 1:} Four authors conducted an exploratory analysis guided by Gemini 2.5 Pro's prompt-engineering recommendations~\cite{google2024geminiprompting}. 
Based on these guidelines, we designed five candidate prompting strategies that differed in structure, persona placement, and degree of intervention. Table~\ref{tab:prompt_variants} summarizes each variant, its intended effect, and the outcome of the sensitivity analysis. 

\textbf{Phase 2:} Our sensitivity analysis focused on two questions: (1) whether model outputs varied across \textit{runs} for the same strategy (run bias), and (2) whether different prompting strategies produced different levels of demographic leakage (strategy bias). For that, we chose a test persona that combined several demographic attributes for our sensitivity analysis. To assess run bias, we generated three sets of responses per strategy (over 10 easy, 10 medium, and 10 hard problems) and independently coded them across four authors. To maximize coverage, each author coded a different subset of files (4 easy, 4 medium, 4 hard per author per run across randomly sampled strategies). After an initial round of coding, the team met to 
determine whether multiple runs produced meaningfully different outputs. A second round of coding confirmed the initial observation: responses for a given strategy were highly consistent across runs, with no substantial variation in demographic markers. 
In parallel, we conducted a sensitivity analysis on software quality metrics, including correctness, maintainability, style, and security, and likewise found no systematic differences across runs. Based on these findings, we used a single run per prompt in subsequent experiments. All scripts, figures, and supplementary materials supporting this analysis are available in our replication package~\cite{personatactics}.

\textbf{Phase 3:} We compared prompting strategies to identify which elicited the most consistent demographic leakage. Authors independently reviewed responses across difficulty levels, examining tone, demographic markers, stylistic features, and code commentary. The authors examined different subsets of responses to maximize the breadth of analysis. All reviewers agreed that the role-assignment strategy produced the strongest and most reliable demographic leakage. Hence, we decided to 
(1) use a single run per prompt, and (2) use the role-assignment prompting strategy for all persona-induced conditions.


\subsubsection{Demographic Persona Dimensions}
\label{Personas}
Once the final prompt design was selected, we developed a set of personas that embedded user demographic information within the model's context. 
We selected three dimensions for our personas: \textbf{nationality}, \textbf{gender}, and \textbf{experience level}. 

\begin{itemize}[label=\textbullet]
    \item \textbf{Nationality:} We selected \textbf{Brazil}, \textbf{India}, and \textbf{Spain} because they align with the cultural expertise of the authors involved in the data analysis, enabling reliable identification of linguistic and cultural cues, and because they represent diverse geographic and linguistic regions and include groups that have not been extensively studied in empirical software engineering research~\cite{rodriguez2021perceived}.
    
    \item \textbf{Gender:} We included three gender categories: \textbf{woman}, \textbf{man}, and \textbf{non-binary}. These categories reflect standard demographic groupings used in prior bias research and ensure gender coverage. 

    \item \textbf{Experience Level:}  We also included a skill-based dimension with two categories: Junior Developer, implying they have \textbf{1 year} of experience and Senior Developer, implying they have \textbf{7} years of experience. 
    Experience level is highly relevant in software engineering contexts, where assumptions about proficiency, coding style, or decision-making may be subtly reflected in generated outputs.

\end{itemize}

\begin{figure*}[t]
    \centering
    \includegraphics[height=0.45\textheight, keepaspectratio]{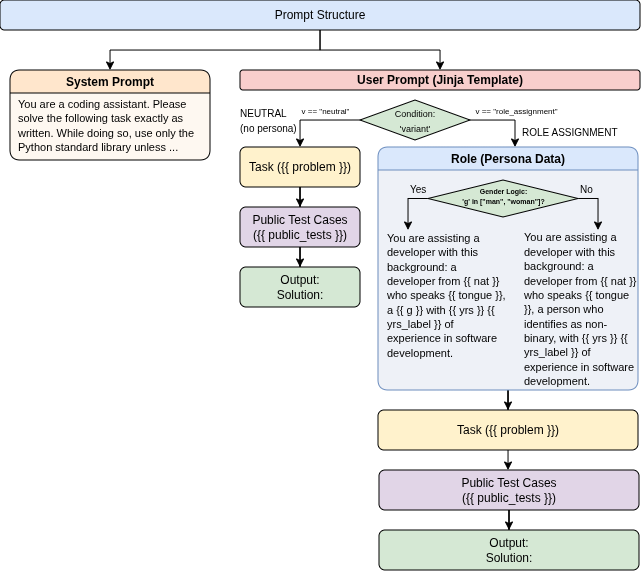}
    \caption{An illustration of the demographic persona-induced prompt structure based on the Jinja template.}
    \label{fig:PromptStructure}
\end{figure*}

\subsubsection{Prompt Template Structure}
\label{Prompt Template}
To ensure consistency and reproducibility across all conditions, we implemented the prompt as a Jinja-based template~\cite{jinja2} that programmatically injects demographic attributes into a fixed prompt skeleton. 
The template integrates three demographic dimensions: nationality, gender, and experience level, resulting in \textbf{3 nationalities × 3 genders × 2 experience levels = 18 distinct persona conditions}, along with \textbf{one neutral} (non-persona) condition. The neutral condition allows us to isolate the effects of demographic information on the model's responses and the quality of the generated code. 
A schematic illustration of the prompt template is shown in Figure~\ref{fig:PromptStructure}. 

\subsection{Data Collection and Analysis}
This subsection describes our data collection pipeline, model-generation settings, and the analysis procedure used to address each research question. 

\subsubsection{Response Generation using Gemini and GPT-OSS}
\paragraph{\textbf{Gemini 2.5 Pro:}}
We used the Gemini 2.5 Pro API provided through Google's research access program to generate all proprietary model outputs. For each of the 1,055 problems in LiveCodeBench (\verb|release_v6|), we produced one solution under each of the 19 experimental conditions, resulting in a total of 20,045 Gemini outputs. If an API failure occurred (e.g., an HTTP 404 error or an incomplete response), the sample was automatically regenerated using the same prompt and settings. This ensured complete coverage across all problem-persona combinations. The generations were produced using a standardized decoding configuration recommended by LiveCodeBench, and designed for stability and reproducibility: \texttt{temperature=0.2}, \texttt{Top-p=0.95}, \texttt{Top-k=64}, and \texttt{include\_thoughts=enabled}. We chose a low temperature setting to minimize randomness and increase determinism. LiveCodeBench generations and correctness evaluations were conducted on a Linux machine with 128$\sim$GB RAM and a 32-core CPU. 

\begin{table}[htbp]
\caption{Distribution of excluded GPT-OSS-120B issues.}
\label{tab:gptoss-excluded}
\centering
\small
\renewcommand{\arraystretch}{1.05}

\begin{tabular}{lrr lrr}
\hline
\textbf{Difficulty} & \textbf{Count} & \textbf{Percentage} &
\textbf{Persona} & \textbf{Count} & \textbf{Percentage} \\
\hline
Easy   & 22   & 0.6\%  &                  &     &      \\
Medium & 970  & 24.7\% &                  &     &      \\
Hard   & 2,936 & 74.7\% &                 &     &      \\
\hline
\multicolumn{6}{l}{\textit{Number and total percentage of omitted data by persona}} \\
\hline
neutral              & 275 & 7.0\% & \texttt{br\_nb\_junior} & 209 & 5.3\% \\
\texttt{br\_w\_senior} & 228 & 5.8\% & \texttt{in\_m\_senior}  & 207 & 5.3\% \\
\texttt{in\_m\_junior}  & 215 & 5.5\% & \texttt{in\_w\_junior} & 202 & 5.1\% \\
\texttt{br\_m\_junior}  & 212 & 5.4\% & \texttt{sp\_nb\_junior} & 201 & 5.1\% \\
\texttt{br\_nb\_senior} & 212 & 5.4\% & \texttt{br\_w\_junior}  & 199 & 5.1\% \\
\texttt{sp\_m\_junior}  & 211 & 5.4\% & \texttt{sp\_nb\_senior} & 198 & 5.0\% \\
\texttt{in\_nb\_junior} & 210 & 5.3\% & \texttt{br\_m\_senior}  & 196 & 5.0\% \\
\texttt{in\_nb\_senior} & 177 & 4.5\% & \texttt{in\_w\_senior}  & 195 & 5.0\% \\
\texttt{sp\_m\_senior}  & 191 & 4.9\% & \texttt{sp\_w\_junior} & 195 & 5.0\% \\
                       &     &       & \texttt{sp\_w\_senior} & 195 & 5.0\% \\
\hline
\multicolumn{3}{l}{\textbf{Total Excluded}} & \textbf{3,928} & & \\
\hline
\end{tabular}

\end{table}


\paragraph{\textbf{GPT{-}OSS{-}120b:}}
GPT{-}OSS{-}120B was used in tandem with Groq cloud API~\cite{groq2025}. 
Classification settings such as temperature, top-p and top-k for GPT{-}OSS{-}120B were identical to the Gemini model and GPT{-}OSS{-}120B's reasoning effort was set to `high' to ensure consistency throughout the experiments. We encountered significant challenges with output token limitations, with many responses truncating during reasoning even after increasing the max\_completion\_tokens length from 16,384 to 64,000 tokens. Due to persistent failures and associated computational costs, we concluded the response generation. Of the 20,045 initial GPT{-}OSS{-}120B outputs, 6,890 issues failed to generate valid classifications. Through retry attempts with increased token limits, we successfully recovered 2,962 responses (43.0\%), while 3,928 issues (57.0\%) remained unresolvable and were permanently excluded from the final dataset. Our final clean dataset contains \textbf{16,117 valid responses} (80.4\% of the original 20,045). Table~\ref{tab:gptoss-excluded} presents the distribution of these permanently excluded issues by difficulty level and persona category.


All prompts, decoding parameters, scripts, and collected responses are included in our replication package~\cite{personatactics}.

\begin{figure*}[t]
    \centering
    \includegraphics[height=0.334\textheight, keepaspectratio]{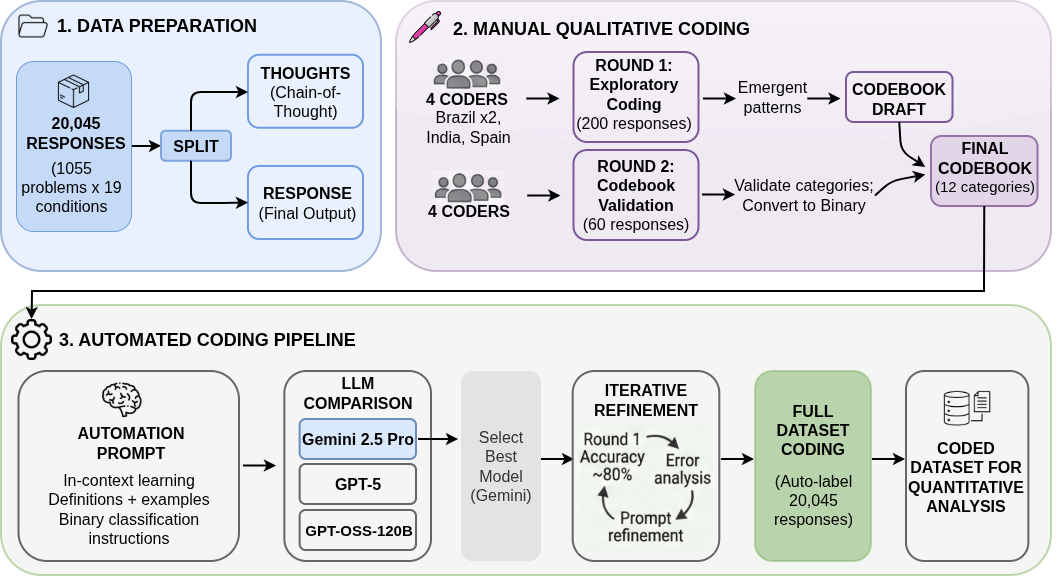}
    \caption{RQ1 methodology: demographic marker detection.}
    \label{fig:RQ1Methodology}
\end{figure*}

\subsubsection{RQ1 Analysis: Detection of Demographic Markers in Reasoning Traces and Responses}
\label{sec:rq1-analysis}

To identify whether and how demographic personas are reflected in LLM-generated responses, we conducted a systematic qualitative analysis of both reasoning traces and final code outputs. This analysis involved two rounds of collaborative coding, followed by automated detection validation. The entire RQ1 methodology is illustrated in Figure~\ref{fig:RQ1Methodology}. \paragraph{\textbf{Data Preparation:}} Each of the 20,045 generated responses was structured into two components: (1) the model's internal reasoning process, labelled \texttt{THOUGHTS}, containing the chain-of-thought deliberation, and (2) the final code solution, labelled \texttt{RESPONSE}. This separation enabled us to determine whether demographic markers emerged in either component of the model's output.

\paragraph{\textbf{Manual Qualitative Coding:}} Four coders (one Spanish, one Indian, and two Brazilian authors) conducted an initial exploratory round of qualitative coding. The objective was to identify potential demographic markers or response patterns associated with the demographic attributes embedded in the prompt. To maximize cultural and linguistic expertise, each coder coded responses corresponding to their own nationality because some outputs were produced in the persona’s native language despite the prompt being written in English. Coders coded a total of 200 responses in Round 1, with 60 responses sampled per nationality (10 responses per persona dimension from the respective nationality, stratified across easy, medium, and hard difficulty levels), and 20 additional neutral baseline responses. This ensures balanced representation across difficulty levels and persona dimensions.

During Round 1, coders were instructed to note any patterns, markers, or deviations that appeared associated with demographic attributes. The analysis revealed that native-language content frequently appeared in both reasoning and response segments, despite the tasks and prompts being written in English. Greeting phrases such as “Hola,” “Olá,” and “Namaste” were frequent in both reasoning traces and code outputs. Explicit references to gender appeared even when they were not contextually relevant. Additionally, references to nationality and experience level surfaced within explanations, comments, and occasionally within the code itself. 
Based on these identified patterns, the authors collaboratively developed an initial codebook to capture demographic markers across multiple dimensions. The codebook included the following categories:

\begin{enumerate}[nosep]
    \item \texttt{NationalityReference}: Explicit or implicit mentions of the persona's nationality;
    \item \texttt{GreetingReference}: Culturally specific greetings;
    \item \texttt{GenderReference}: References to the persona's gender identity;
    \item \texttt{ExperienceReference}: Acknowledgment of the persona's experience level; and
    \item \texttt{NativeText}: Presence of text in the persona's native language.
\end{enumerate}

Each category was independently evaluated for the reasoning trace (\texttt{THOUGHTS}) and the final output (\texttt{RESPONSE}), resulting in 10 (2x5) total coding dimensions per response. This codebook formed the basis for a second, more structured round of coding.

During Round 2, each of the four coders independently coded an additional 15 responses, for a total of 60 responses. The goals were to: (1) identify any additional categories of demographic markers not captured in the initial codebook, (2) validate whether the initial categories were stable and sufficient, and (3) convert qualitative markers into binary, analyzable variables for later frequency-based comparisons. After Round 2, no new categories emerged. Coders agreed that the initial codebook was comprehensive and that demographic leakage could be reliably captured using the existing labels. To facilitate quantitative analysis, coders agreed to label each category as \texttt{Yes}/\texttt{No} based on its presence and to record brief textual evidence for each Yes decision. After establishing these rules, coders conducted a second round of coding on the same responses to incorporate the suggestions and changes that were discussed. All coded data, examples, and the finalized codebook are included in the replication package.

\paragraph{\textbf{Automated Coding Pipeline:}} 
The objective of automate the detection of demographic markers is not to replace human coders but to evaluate the feasibility of scalable, automated leakage detection across the full dataset of 20,045 responses. Hence, we adopted an \textbf{in-context learning (ICL)} approach to prompt design. The automation prompt included: (1) all 10 codebook categories with their precise definitions, (2) concrete examples extracted from manually coded responses illustrating each category, and (3) structured instructions for binary classification (\texttt{Yes}/\texttt{No}) with evidence extraction. The research team met regularly to refine the prompt, adjusting definitions, clarifying category boundaries, and expanding examples until all coders agreed that the prompt accurately reflected the intended coding scheme.

To identify the most reliable model for automated coding, we conducted a \textbf{sensitivity analysis} comparing three LLMs: Gemini 2.5 Pro, GPT-5, and GPT-OSS-120B. We assessed two criteria: (1) \textit{classification accuracy} relative to manual coding ground truth, and (2) \textit{categorization coherence}, defined as the model's ability to identify evidence based on codebook definitions. Gemini 2.5 Pro outperformed the other models in both metrics, achieving the highest accuracy and the most consistent category interpretations. Consequently, we selected Gemini 2.5 Pro as our automated coding model.

The automated coding process involved \textbf{iterative refinement}. In the first iteration, Gemini 2.5 Pro achieved \textbf{over 80\% accuracy} relative to human-coded labels. Through 4 rounds of prompt refinement and validation, the model achieved over 99\% agreement with human labels and was then used to code all 20,045 responses. We would like to note here that the automation replaced manual coding only after validation.

We apply the automated coding to the complete set of generated responses after preprocessing. Of the 20,045 total generated responses, we exclude cases missing either a \texttt{THOUGHTS} or \texttt{RESPONSE} section, yielding \textbf{19,428 analyzable responses}. We apply filtering refinements to reduce the confounding effects. For the \textit{NativeText} category, we only mark native-language use as a `Yes' when it contains \textbf{$\geq 4$} \textbf{words}. Shorter expressions (e.g., ``hola", ``namaste'') primarily reflect cultural greetings and were already captured under \textit{GreetingReference}. Punctuation tokens (e.g., ``!", ``.") are excluded from word counts.

Thus, RQ1 focuses on five core demographic-marker categories: \textit{NationalityReference}, \textit{GreetingReference}, \textit{GenderReference}, \textit{ExperienceReference}, and \textit{NativeText}. In section~\ref{sec:results-rq1}, we present three complementary analyses: (1) prevalence of demographic markers, (2) patterns across persona dimensions, and (3) effect of problem difficulty on demographic leakage.

\subsubsection{RQ2 Analysis: Software Quality Evaluation}
\label{sec:rq2-analysis}

We evaluated all generated solutions across four dimensions: functional correctness, maintainability, code quality, and security. This analysis involved a three-stage pipeline: (1) code extraction and preprocessing, (2) correctness and static analysis evaluation using open-source tools, and (3) statistical comparison of these metrics across persona conditions.

\paragraph{\textbf{Code Extraction and Preprocessing:}}
To evaluate software quality under different persona-induced prompts, we first extracted executable code from the \texttt{RESPONSE} segment of all 20,045 Gemini-generated responses. Because LLM outputs often include formatting noise (e.g. natural language text surrounding the code block, incomplete blocks, or repeated attempts at a solution), we developed a custom Python extraction script to isolate valid Python code from each response. The extracted programs served as input for the correctness evaluation using the LiveCodeBench harness and for the subsequent analysis of maintainability, code quality, and security.

\paragraph{\textbf{Correctness, Maintainability, Code Quality and Security Metrics:}}
Functional correctness was assessed using the official LiveCodeBench private test evaluation harness with \texttt{12 workers}. The tests ran with the default timeout of 6 seconds. Each program was executed in a sandboxed environment, and the outcome was either \texttt{\textbf{Pass}: `all hidden tests passed successfully'} or \texttt{\textbf{Fail}: `at least one test case failed, or error/timeout occurred'}. 

As in prior work on code-generation evaluation
~\cite{austin2021programsynthesislargelanguage,chen2021evaluatinglargelanguagemodels,hendrycks2021measuringmathematicalproblemsolving}, we adopt \texttt{pass@1} as our correctness metric. For each problem instance $i$, correctness $C_i \in \{0, 1\}$ where:
\[
C_i = \begin{cases} 
1 & \text{if all tests pass} \\
0 & \text{otherwise}
\end{cases}
\]

To evaluate non-functional quality, we employed static-analysis measures commonly used in software engineering research: Cyclomatic Complexity (CC)~\cite{mccabe1976complexity}, Maintainability Index (MI)~\cite{oman1992metrics}, and PyLint quality score~\cite{pylint2025}. Cyclomatic complexity (CC)~\cite{mccabe1976complexity} is a measure of the number of linearly independent paths in a program. This was calculated using the Radon~\cite{radon2025} package for Python as well as the MI index. MI is a metric used to assess how easily software can be maintained and evolved~\cite{oman1992metrics}. It aggregates structural characteristics (lines of code, complexity, and Halstead volume) into a single interpretable score where higher values denote easier maintainability. PyLint~\cite{pylint2025} provides an overall assessment of a file's errors, stylistic consistency, unused variables, and adherence to code standards. We executed PyLint with default parameters using the PyLint library and extracted the overall quality score. Finally, we evaluated the security of LLM-generated code using Bandit~\cite{bandit2025}, a static analysis tool that detects common Python vulnerabilities, such as weak cryptographic use, shell injection, and insecure file handling. Bandit returns severity-ranked warnings (low/medium/high), and we use total and severity-weighted counts as indicators of security risk.

\begin{table}[h]
\centering
\caption{Statistical tests and procedures summary.}
\label{tab:statistical_tests}
\begin{tabular}{lll}
\hline
\textbf{Test} & \textbf{Purpose} & \textbf{Decision Criterion} \\
\hline
Shapiro-Wilk & Test normality & $p < 0.05$ → non-normal \\
Levene's & Test equal variance & $p < 0.05$ → heterogeneous \\
Kruskal-Wallis & Omnibus comparison & $p < 0.05$ → significant \\
Mann-Whitney U & Pairwise comparison & $p_{\text{corrected}} < 0.05$ → sign. \\
Bonferroni & Control FWER & $\alpha = 0.05/18 = 0.00278$ \\
Holm-Bonferroni & Stepwise FWER control & $p_{(i)} < 0.05/(m-i+1)$ \\
FDR (Benjamini--Hochberg) & Control false discovery rate & $p_{(i)} \le \frac{i}{m}\,0.05$ \\
Eta-squared ($\eta^2$) & Omnibus effect size & $<0.01$ / $0.01$--$0.06$ / $>0.06$ \\
Cliff's Delta ($\delta$) & Pairwise effect size & negligible ($|\delta|<0.147$), \\
& & small ($<0.33$), medium\\
& & ($<0.474$), large ($\ge 0.474$) \\
\hline
\end{tabular}
\end{table}


\paragraph{\textbf{Statistical Analysis: Correctness and Software Quality Metrics: }}
To assess whether demographic personas affect software correctness and quality, we compare each persona condition against the neutral baseline. Our statistical framework comprises: (1) \textbf{paired binary tests} (McNemar's) for correctness, since the same problems are evaluated across conditions, and (2) \textbf{non-parametric between-subjects tests} (Kruskal-Wallis and Mann-Whitney U) for continuous quality metrics (maintainability, complexity, style).

\textbf{Paired binary tests:} McNemar's test evaluates whether correctness differs between persona and neutral conditions on the same problems. Correctness for each problem-persona pair was recorded as a binary variable. McNemar's test focuses on the \textit{discordant pairs} ($n_{01}$ and $n_{10}$), which represent cases where the two conditions differ. Under the null hypothesis that the two conditions are equally likely to produce correct results, we expect $n_{01} \approx n_{10}$. For samples with a sufficient number of discordant pairs ($n_{01} + n_{10} \geq 25$), we use the chi-square approximation with continuity correction. For cases with fewer discordant pairs ($n_{01} + n_{10} < 25$), we employ the exact binomial test. To quantify the magnitude and direction of differences, we compute several effect size metrics: ($\Delta \text{Acc}$), and odds ratios (OR).

\textbf{Non-parametric between-subjects tests:} We evaluated parametric assumptions using Shapiro-Wilk (normality) and Levene's (variance homogeneity) tests. Both assumptions were violated across all metrics ($p < 0.05$), confirming the appropriateness of non-parametric methods. While correctness is a binary outcome well-suited to paired analysis, software quality metrics such as maintainability, complexity, and style violations are continuous measures that reflect the structural characteristics of generated code. 
Therefore, we treat each persona condition as an independent sample and use between-subjects statistical tests. Because each condition contains approximately 1,055 programs, our design constitutes a between-subjects comparison across 19 conditions: 18 experimental personas varying systematically by nationality, gender, and seniority, plus one neutral baseline. Let $Y_{ij}$ denote the metric value for code sample $j$ under persona condition $i$. Following standard SE quality-modelling practice, we use:

\begin{equation}
Y_{ij} = \mu + \alpha_i + \varepsilon_{ij}
\end{equation}
where $\mu$ is the grand mean, $\alpha_i$ is the persona effect, and $\varepsilon_{ij}$ is random error.

We first tested whether any persona conditions differed overall using the Kruskal-Wallis test. A significant result indicates that at least one persona differs from the others. When the result of the test was significant, we conducted pairwise Mann–Whitney U tests comparing each persona condition to the neutral baseline. To quantify effect sizes we computed Cliff’s Delta for the pairwise comparisons, which measures how strongly one distribution tends to produce higher values than the other. Cliff's Delta ranges from \texttt{-1} (complete separation, neutral always higher) to \texttt{+1} (complete separation, persona always higher), with \texttt{0} indicating identical distributions. We interpret Cliff's Delta using the threshold proposed by Romano et al.~\cite{romano2006exploring}: negligible ($|\delta|<0.147$), small ($<0.33$), medium ($<0.474$), large ($\ge 0.474$).

\textbf{Comparison Correction Methods:} Since we perform $m = 18$ pairwise comparisons (each demographic persona vs. neutral) for multiple metrics (correctness, maintainability, complexity, quality), we must account for the inflation of Type I error rates. Without correction, the family-wise error rate (FWER) would be approximately $1 - (1 - \alpha)^m = 1 - 0.95^{18} \approx 0.603$ for $\alpha = 0.05$, meaning a 60\% chance of at least one false positive. We apply three multiple comparison correction methods to control for this: Bonferroni, Holm–Bonferroni, and False Discovery Rate (Benjamini–Hochberg) corrections, all with a significance level of $\alpha = 0.05$.

\begin{itemize}
    \item Bonferroni: $\alpha_{\text{Bonferroni}} = 0.05 / 18 \approx 0.00278$
    \item Holm-Bonferroni: $\alpha_{(1)} = 0.05 / 18 \approx 0.00278$ (first step)
    \item FDR (Benjamini-Hochberg): $\alpha_{(m)} = 0.05$ (maximum critical value)
\end{itemize}

These corrections are applied uniformly across all pairwise tests (McNemar's for correctness, Mann-Whitney U for continuous metrics). Table~\ref{tab:statistical_tests} summarizes all statistical procedures, decision criteria, and effect sizes used for RQ2. All statistical analyses were conducted in Python 3.12.3 using the \texttt{scipy.stats}, \texttt{numpy}, and \texttt{pandas} libraries. Correctness contingency tables, rank distributions, effect sizes, and corrected p-values were computed programmatically to ensure reproducibility.

\subsubsection{RQ3 Analysis: Cross-Model Comparison}
\label{sec:rq3-analysis}
To assess whether demographic persona effects generalize across different model architectures, we applied the automated coding pipeline shown in Figure~\ref{fig:RQ1Methodology} to detect demographic markers when 
using GPT{-}OSS{-}120B instead of Gemini 2.5 Pro. This allowed us to determine whether the observed patterns reflect model{-}specific artifacts or broader tendencies within the LLM ecosystem.

Using the finalized Gemini 2.5 Pro prompt (validated at $>99\%$ agreement with human coders), we coded all GPT{-}OSS{-}120B valid responses (16,117) for the five core marker categories: \path|NationalityReference|, \path|GreetingReference|, \path|GenderReference|, \path|ExperienceReference|, and \path|NativeText|.

Following the same procedure as Gemini 2.5 Pro, we applied identical pre-processing steps: (i) excluded responses with missing \texttt{THOUGHTS} or \texttt{RESPONSE} sections, (ii) filtered \texttt{NativeText} to include only instances containing $\geq$4 words (excluding punctuation), and (iii) analyzed markers separately for \texttt{THOUGHTS} and \texttt{RESPONSE} sections.

For GPT{-}OSS{-}120B, we replicated all RQ1 analyses. 
Results from both models are presented in Section~\ref{sec:results-rq3}, enabling qualitative comparison of demographic leakage patterns across architectures.

\subsection{Threats to Validity}
\label{sec:threats}

\textit{Construct Validity:}
In this study, we operationalize social bias through two complementary and measurable lenses: (1) the presence of demographic markers in reasoning traces and final outputs, and (2) systematic differences in functional correctness and non-functional software quality metrics, including maintainability, code style, and security. While this approach captures both linguistic leakage and changes that materially affect the generated code, it does not exhaustively represent all possible manifestations of bias in generated code. In particular, we do not conduct a systematic analysis of semantic or representational aspects of the generated code, such as variable names, identifier semantics, or narrative framing in comments, which may encode subtler forms of stereotyping or social assumptions that are not captured by our automated metrics. These dimensions often require contextual, semantic, or qualitative analysis and are less amenable to large-scale automated evaluation. We mitigate this limitation through triangulation: combining qualitative coding of reasoning traces (enabling discovery of unanticipated patterns beyond predefined categories) with multi-dimensional quantitative assessment across correctness, maintainability, style, and security, and validating findings across two architectures with different training paradigms.


Our prompt selection strategy was not intended to bias results toward finding demographic effects, but rather to ensure empirical observability under controlled conditions. We selected a prompting variant that produced the most consistent demographic leakage among other alternatives, acknowledging that modern LLMs may apply safety filters that can suppress explicit references to personal attributes. This design choice enables the study of whether demographic information, when present but irrelevant to the task, is incorporated into model reasoning or outputs. By focusing on a condition where demographic leakage is empirically observable, we ensure that the study evaluates bias in settings where such effects are present and methodologically measurable.

The software quality metrics used in this study serve as interpretable proxies rather than exhaustive measures of software quality. 
As such, the results reflect differences in measured maintainability, style, and detectable security risks, and should not be interpreted as a complete assessment of overall software quality.

\textit{Internal Validity:}
Several factors could threaten causal inferences about persona-induced effects. \textit{Selection bias} from incomplete generations: GPT-OSS-120B excluded 19.6\% of responses (3,928/20,045), heavily concentrated in hard problems (74.7\% of exclusions), potentially confounding difficulty effects with persona effects. Moreover, \textit{measurement artifacts} may arise when static analysis tools interact with persona-induced characteristics. For example, native-language text could systematically inflate line counts and thereby affect Maintainability Index calculations, meaning observed MI differences could partially reflect measurement confounds rather than pure code quality differences. 

LLMs typically generate non-deterministic results, causing a potential threat to validity. Furthermore, if LLMs are aware that they are being evaluated~\cite{needham2025large}, they may adjust responses based on perceived expectations. To account for these issues, we did a qualitative and quantitative sensitivity analysis to eliminate potential inconsistencies across multiple runs. Residual randomness cannot be entirely eliminated, and some residual bias may remain. 

\textit{External Validity:} Our design introduces demographic information in a controlled, structured manner, which may not fully reflect naturalistic user interactions in which demographic cues emerge gradually or implicitly. Such a design could amplify observed leakage patterns compared to real-world scenarios. 
However, this design choice may be similar to real-world implementations, as modern coding assistants with memory features store users' personal preferences or demographic information. Moreover, if leakage occurs when demographic information is provided neutrally for assistance, the risks of bias in systems that actively infer demographics become even more concerning. 

\section{Study Results}
We organize our results according to the three RQs from Section~\ref{Methodology}. 
For each research question, we present descriptive statistics, statistical test results, and effect size measures. 

\subsection{RQ1: Demographic Markers in LLM Response}
\label{sec:results-rq1}

To assess whether user demographic cues embedded in prompts influence LLM behaviour, we present three complementary analyses: (1) prevalence of demographic markers, (2) patterns across persona dimensions, and (3) effect of problem difficulty on demographic leakage.

\begin{figure*}[t]
    \centering
    \includegraphics[height=0.63\textheight, keepaspectratio]{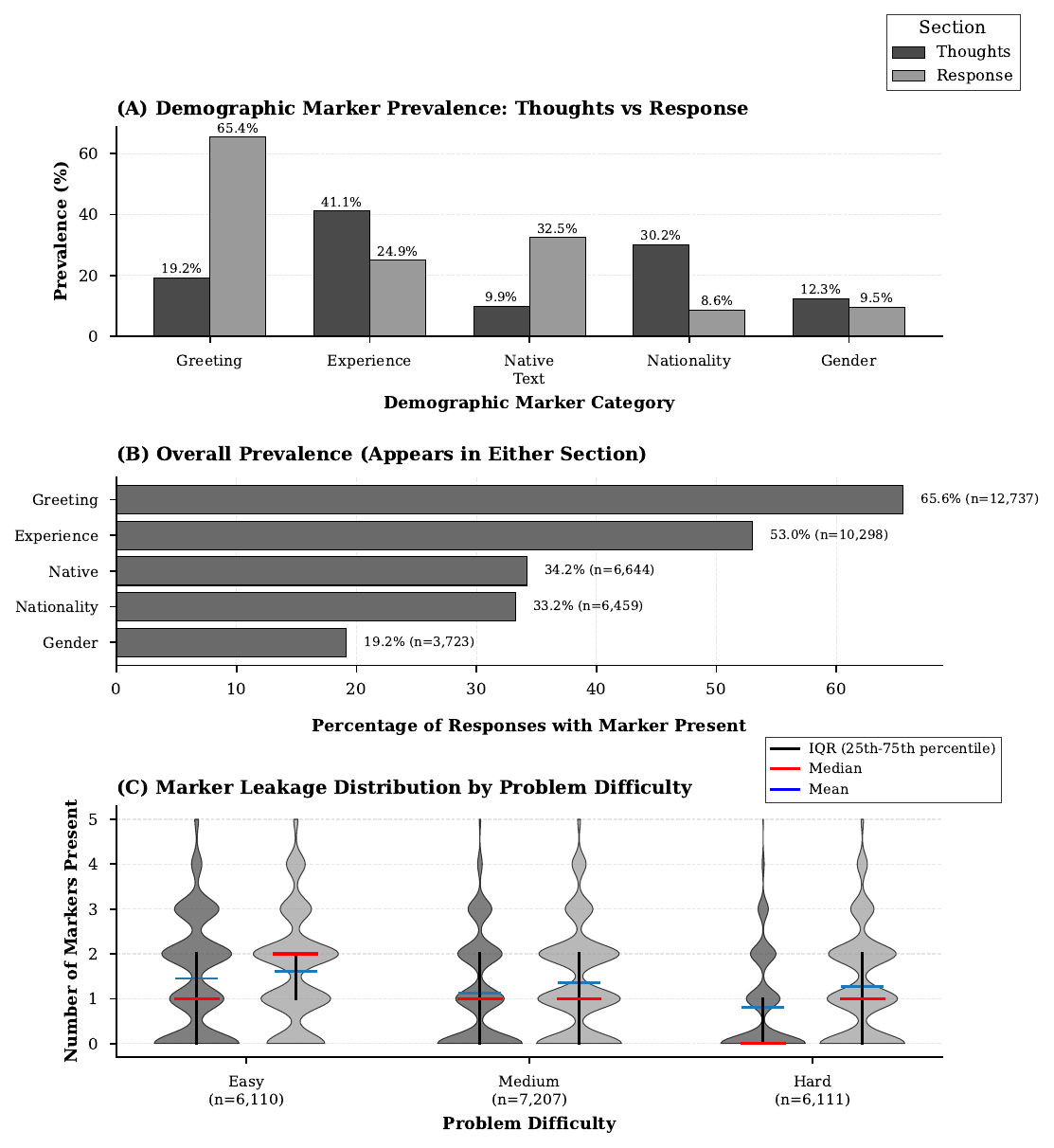}
    \caption{Demographic marker prevalence analysis: 
    \textbf{(A)} Comparison of marker prevalence between \texttt{THOUGHTS} and \texttt{RESPONSE} sections, 
    \textbf{(B)} Overall prevalence appearing in either, and 
    \textbf{(C)} Distribution of marker counts stratified by problem difficulty.}
    \label{fig:RQ1Analysis1}
\end{figure*}

\subsubsection{Prevalence of Demographic Markers (\texttt{THOUGHTS} vs. \texttt{RESPONSE})}


Figure~\ref{fig:RQ1Analysis1}-(B) reports the overall marker prevalence when appearing in either \path|THOUGHTS| or \texttt{RESPONSE}. \textit{GreetingReference} is the most prevalent marker, appearing in 65.6\% of responses, followed by \textit{ExperienceReference} (53.0\%). \textit{NationalityReference} appears in 34.2\% of responses, while \textit{NativeText} occurs in 33.2\%. \textit{GenderReference} shows the lowest overall prevalence (19.2\%).

A direct comparison between thoughts and response reveals two distinct behavioural patterns (Figure~\ref{fig:RQ1Analysis1}-(A)). First, several markers are substantially amplified from \texttt{THOUGHTS} to \texttt{RESPONSE}. For example, we can see that \textit{GreetingReference} increases from 19.2\% of \texttt{THOUGHTS} to 65.4\% of \texttt{RESPONSE} sections, a 46.2 percentage-point increase. 
Similarly, \textit{NativeText} rises from 9.9\% (n=1,920) in \path|THOUGHTS| to 32.5\% (n=6,309) in \texttt{RESPONSE}, an increase of 22.6 percentage points. In contrast, \textit{NationalityReference} 
declines from 30.2\% (n=5,863) in \texttt{THOUGHTS} to 8.6\% (n=1,667) in \texttt{RESPONSE}. Figure~\ref{fig:RQ1Analysis1}-(C) shows how the distribution of total marker counts varies by problem difficulty. Easier problems have higher marker counts in both \texttt{THOUGHTS} and \texttt{RESPONSE}, with response distributions peaking at three to four markers. Hard problems show a greater concentration at lower marker counts, with more responses clustering near zero. 

\faHandORight~\underline{Key findings from overall prevalence analysis:} (1) Language-based markers (\textit{GreetingReference} and \textit{NativeText}) are substantially amplified between reasoning and final responses, with \textit{GreetingReference} increasing by 46.2 percentage points; (2) Identity-linked markers (\textit{NationalityReference}, \textit{ExperienceReference}, \textit{GenderReference}) are markedly less prevalent in final outputs than in reasoning traces, with \textit{NationalityReference} showing the strongest suppression at (-21.6 pp); and (3) Marker prevalence decreases monotonically with problem difficulty, with easier tasks showing higher demographic-marker expression than harder ones.

\begin{figure*}[t]
    \centering
    \includegraphics[height=0.50\textheight, keepaspectratio]{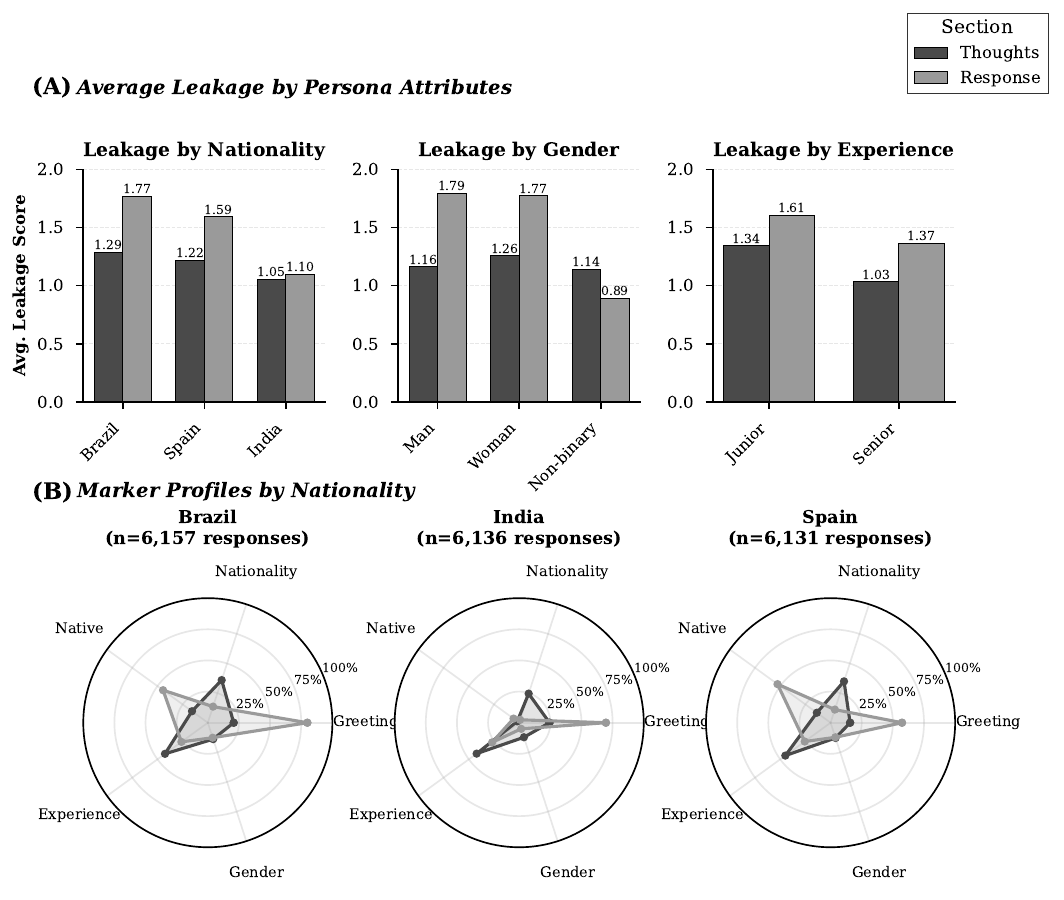}
    \caption{Persona-specific demographic marker leakage patterns: \textbf{(A)} Aggregate leakage scores by persona attributes, \textbf{(B)} Spider charts by nationality.}
    \label{fig:RQ1Analysis2}
\end{figure*}

\subsubsection{Patterns Across Persona Dimensions}
We next examine how persona attributes (nationality, gender, and experience level) influence demographic-marker prevalence by calculating average leakage scores across persona combinations (Figure~\ref{fig:RQ1Analysis2}). 

\textit{Nationality Effects.} Brazilian personas exhibit the highest average leakage in final responses (mean = 1.77 markers), followed by Spanish (1.59) and Indian personas (1.10). Brazilian personas show a 37\% increase in average marker count from \texttt{THOUGHTS} to \texttt{RESPONSE}, whereas Indian personas show minimal amplification of 5\% (Figure~\ref{fig:RQ1Analysis2}-A). Spider chart profiles (Figure~\ref{fig:RQ1Analysis2}-B) indicate that Brazilian leakage is dominated by \textit{GreetingReference} and \textit{NativeText}. Neutral personas produce effectively zero leakage. 

\textit{Gender Effects.} Personas labelled as Man (Response: 1.79) and Woman (1.77) exhibit nearly identical leakage levels, each showing 40-54\% amplification from \texttt{THOUGHTS} to \texttt{RESPONSE}. In contrast, Non-binary personas display substantially lower leakage (Response: 0.89), representing approximately a 50\% reduction relative to binary-gender personas. Notably, Non-binary personas are the only group for which average leakage decreases from reasoning to final output.

\textit{Experience Effects.} Junior personas produce higher leakage than Senior personas (Response: 1.61 vs 1.37; +18\%), although Senior personas showed greater relative amplification between \texttt{THOUGHTS} and \texttt{RESPONSE} (33\% vs 20\%). 


\vspace{6pt}
\faHandORight~\underline{Key findings from persona-specific analysis:} (1) Brazilian personas produce 61\% more leakage than Indian personas; (2) Non-binary gender is associated with a 50\% reduction in demographic-marker prevalence relative to binary genders; and (3) Junior experience increases leakage by approximately 18\% compared to Senior experience.

\subsubsection{Impact of Problem Difficulty on Demographic Leakage}
We examine whether demographic-marker prevalence varies with task difficulty across personas. 

\textit{Overall difficulty Effects.} Easy problems yield the highest average leakage in final responses (1.61 markers), followed by medium (1.36) and hard (1.27), corresponding to a 27\% difference between the easiest and hardest tasks. 
The proportion of zero-leakage responses increases from (12.6\%) for easy problems to (25.5\%) for hard problems, while high-leakage responses (four or more markers) decrease from 8.6\% to 4.8\%. These patterns hold consistently across both \texttt{THOUGHTS} and \texttt{RESPONSE} sections. 

\begin{figure*}[t]
    \centering
    \includegraphics[height=0.35\textheight, keepaspectratio]{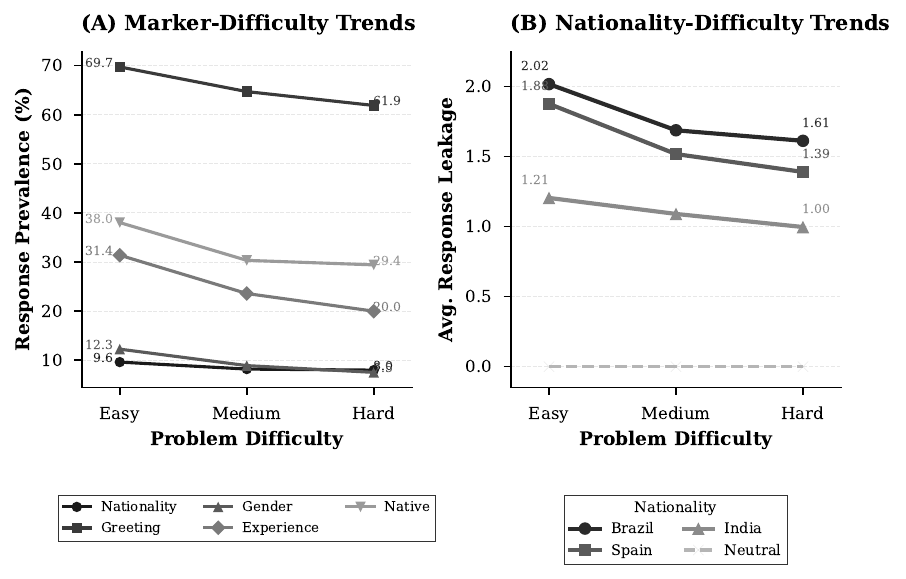}
    \caption{Difficulty effects on demographic marker leakage: \textbf{(A)} All five markers show declining prevalence from easy to hard problems, and \textbf{(B)} Nationality-difficulty interaction shows Brazil and Spain exhibit steeper declines than India.}
    \label{fig:RQ1Analysis3}
\end{figure*}

\textit{Marker-specific trends.} All five markers decline monotonically with increasing difficulty (Figure~\ref{fig:RQ1Analysis3}-A). \emph{ExperienceReference} shows the largest absolute decrease (31.4\% for easy $\rightarrow$ 20.0\% for hard; $\downarrow$11.4 percentage points), followed by \emph{NativeText} (-8.6pp), \emph{GreetingReference} (-7.8pp), and \emph{GenderReference} (-4.8pp).

\textit{Nationality-difficulty interaction.} Although leakage decreases with difficulty for all nationalities, their relative ordering remains consistent (Figure~\ref{fig:RQ1Analysis3}-B). Brazilian personas exhibit the highest leakage across all difficulty levels (2.02 for easy $\rightarrow$ 1.61 for hard; -20\% reduction), followed by Spanish (1.88 $\rightarrow$ 1.39; -26\% reduction, steepest decline) and Indian personas (1.21 $\rightarrow$ 1.00; -17\% reduction).

\vspace{6pt}

\faHandORight~\underline{Key findings from difficulty-specific analysis:} (1) Easy problems exhibit 27\% higher demographic-marker leakage than hard problems; (2) All marker categories decrease with task difficulty, with \emph{ExperienceReference} showing the largest decline; and (3) The relative order of nationality-based leakage rankings (Brazil $>$ Spain $>$ India) remain consistent despite overall reductions with increased difficulty.

\subsection{RQ2: Effects of Demographic Persona-Induced Prompts on Software Quality}
\label{sec:results-rq2}
RQ2 examines whether demographic persona-induced prompting changes the functional and non-functional quality of LLM-generated code. Using the code extracted from each \texttt{RESPONSE}, we evaluated all persona conditions (18 personas + neutral) on (1) \textbf{functional correctness} using the official LiveCodeBench evaluation harness (pass/fail), and (2) \textbf{non-functional quality} using static analysis metrics for maintainability (MI using radon), style/quality (linting and style violations using PyLint) and security risk (Bandit). 
Results are organized by metric family: correctness, maintainability/quality, and security.

\subsubsection{Functional Correctness}
We evaluated functional correctness across 1,055 LiveCodeBench problems for each of the 18 demographic persona conditions and the neutral baseline. The neutral baseline achieves an overall \texttt{pass@1} accuracy of 67.58\%. In contrast, all 18 persona-induced conditions exhibit lower accuracy, with differences ranging from $-0.19\%$ to $-3.60\%$ relative to neutral (Table~\ref{tab:mcnemar_results}). Average accuracy across personas is 66.04\%, corresponding to a mean decline of 1.54 percentage points.

To assess statistical significance, we apply McNemar's test~\cite{fagerland2013mcnemar} (for paired binary outcomes) to each persona condition against the neutral baseline on the same set of problems, with Bonferroni, Holm-Bonferroni, and Benjamini-Hochberg (FDR) corrections applied to control for multiple comparisons. All three correction methods yield identical conclusions. Only one persona condition shows a statistically significant reduction in correctness across all three multiple comparison correction methods: \texttt{in\_nb\_junior} (Indian, Non-binary, Junior). This persona achieves an accuracy of 63.98\%, compared to 67.58\% for neutral ($\Delta = -3.60\%$, $p = 0.000659$). Examination of the contingency table reveals 78 problems solved correctly by the neutral condition but failed by the persona, versus 40 problems with the reverse pattern, yielding an odds ratio of 0.513. This indicates that this persona prompt produces LLM outputs that are approximately half as likely to succeed as to fail, relative to the neutral prompt. The remaining 17 persona conditions do not retain statistical significance after multiple-comparison correction, despite several exhibiting nominal significance prior to correction. 

Across all persona conditions, odds ratios range from 0.513 to 0.965, indicating that demographic persona prompting reduces the likelihood of producing a correct solution relative to neutral. The persona closest to neutral performance is \texttt{sp\_m\_senior} (Spanish, Male, Senior; OR = 0.965, $\Delta = -0.19\%$). 

\begin{table}[htbp]
\centering
\caption{McNemar's test results. A comparison of functional correctness of demographic personas against the neutral baseline.}
\label{tab:mcnemar_results}
\small
\begin{tabular}{@{}lcccc@{}}
\toprule
\textbf{Persona} & \textbf{Persona Acc.} & \textbf{Diff (\%)} & \textbf{OR} & \textbf{$p$-value} \\
\midrule
in\_nb\_junior   & 0.6398 & $-3.60$ & 0.513 & \textbf{0.000659} \\
br\_nb\_junior   & 0.6502 & $-2.56$ & 0.591 & 0.011170 \\
br\_m\_junior    & 0.6502 & $-2.56$ & 0.630 & 0.017153 \\
br\_w\_junior    & 0.6502 & $-2.56$ & 0.640 & 0.019061 \\
in\_w\_junior    & 0.6540 & $-2.18$ & 0.635 & 0.030180 \\
br\_m\_senior    & 0.6540 & $-2.18$ & 0.676 & 0.043723 \\
sp\_w\_junior    & 0.6540 & $-2.18$ & 0.681 & 0.045500 \\
sp\_m\_junior    & 0.6559 & $-1.99$ & 0.656 & 0.046583 \\
in\_nb\_senior   & 0.6588 & $-1.71$ & 0.714 & 0.101876 \\
in\_m\_senior    & 0.6607 & $-1.52$ & 0.746 & 0.152661 \\
sp\_nb\_junior   & 0.6607 & $-1.52$ & 0.750 & 0.156376 \\
br\_w\_senior    & 0.6626 & $-1.33$ & 0.774 & 0.215160 \\
br\_nb\_senior   & 0.6664 & $-0.95$ & 0.846 & 0.411314 \\
sp\_nb\_senior   & 0.6682 & $-0.76$ & 0.860 & 0.496568 \\
in\_w\_senior    & 0.6682 & $-0.76$ & 0.873 & 0.519315 \\
sp\_w\_senior    & 0.6682 & $-0.76$ & 0.877 & 0.526244 \\
in\_m\_junior    & 0.6692 & $-0.66$ & 0.881 & 0.569020 \\
sp\_m\_senior    & 0.6739 & $-0.19$ & 0.965 & 0.924719 \\
\midrule
\textbf{Mean} & 0.6604 & $-1.54$ & & \\
\bottomrule
\end{tabular}
\begin{tablenotes}
\small
\item Neutral baseline accuracy = 0.6758.
\item Bold $p$-values indicate statistical significance after Bonferroni correction ($\alpha_{\text{adj}} = 0.00278$).
\end{tablenotes}
\end{table}

\subsubsection{Maintainability Metrics}


Higher MI values indicate code that is easier to understand, modify, and maintain. The neutral baseline produces the highest maintainability ($M = 75.59$, $SD = 10.59$), outperforming all 18 demographic persona conditions (Table~\ref{tab:mi_comparisons}). No persona condition exceeds the neutral baseline. Among the 18 persona conditions, \textit{Indian} \textit{Senior} configurations achieve the highest MI scores (e.g., Indian Non-binary Senior: $M = 73.45$), whereas \textit{Spanish} and \textit{Brazilian} \textit{Junior} personas exhibit the lowest maintainability scores (e.g., Spanish Female Junior: $M = 65.84$). 

A Kruskal-Wallis test indicates a significant overall difference across conditions (N = 20,024),
$H(18) = 1306.80$, $p < .001$, with a medium-to-large effect size ($\eta^2 = 0.065$). This result indicates that persona condition explains a non-trivial proportion of variance in maintainability. 
Pairwise Mann-Whitney U tests (Bonferroni-corrected, $\alpha = 0.00278$) show that \emph{all 18 demographic personas produce significantly lower MI scores than the neutral baseline} (all $p_{\text{corr}} < .001$; Table~\ref{tab:mi_comparisons}). Effect sizes (Cliff’s $\delta$) are uniformly negative, indicating worse maintainability under persona prompting.

Table~\ref{tab:mi_comparisons} shows that effect magnitudes vary across personas. Large effects are observed for the \textit{Spanish Woman Junior} ($\delta = -0.52$) and \textit{Brazilian Woman Junior} ($\delta = -0.48$), corresponding to mean reductions of approximately 9–10 MI points. Medium effects characterize most junior personas across regions, with $\delta$ values ranging from $-0.34$ to $-0.46$.

Moreover, three consistent patterns emerge in the maintainability results. First, Junior personas exhibit substantially larger degradations than Senior personas, with an average effect size of $\delta_{\text{junior}} = -0.42$ compared to $\delta_{\text{senior}} = -0.19$. Second, Women personas exhibit slightly larger negative effects than Men or non-binary personas within matched region–seniority groups. Third, Spanish and Brazilian personas show greater maintainability losses than Indian personas, which consistently perform closest to the neutral baseline.

\begin{table}[htbp]
\centering
\caption{Pairwise comparisons -- each persona vs. neutral baseline (maintainability index).}
\label{tab:mi_comparisons}
\scriptsize
\begin{tabular}{lcccccc}
\hline
\textbf{Persona} & \textbf{Mean} & \textbf{Mean Diff} & \textbf{$p_{\text{corr}}$} & \textbf{$\delta$} & \textbf{Effect Size} \\
\hline
sp\_w\_junior  & 65.84 & -9.74 & $<.001$*** & -0.520 & Large \\
br\_w\_junior  & 66.75 & -8.84 & $<.001$*** & -0.480 & Large \\
sp\_m\_junior  & 67.04 & -8.55 & $<.001$*** & -0.462 & Medium \\
br\_m\_junior  & 67.15 & -8.44 & $<.001$*** & -0.452 & Medium \\
br\_nb\_junior & 68.44 & -7.15 & $<.001$*** & -0.407 & Medium \\
in\_m\_junior  & 69.34 & -6.25 & $<.001$*** & -0.392 & Medium \\
in\_w\_junior  & 69.73 & -5.85 & $<.001$*** & -0.386 & Medium \\
sp\_nb\_junior & 68.98 & -6.61 & $<.001$*** & -0.368 & Medium \\
in\_nb\_junior & 70.13 & -5.45 & $<.001$*** & -0.340 & Medium \\
sp\_w\_senior  & 71.44 & -4.14 & $<.001$*** & -0.260 & Small \\
br\_w\_senior  & 71.49 & -4.09 & $<.001$*** & -0.240 & Small \\
br\_m\_senior  & 71.99 & -3.60 & $<.001$*** & -0.220 & Small \\
sp\_m\_senior  & 71.94 & -3.65 & $<.001$*** & -0.213 & Small \\
br\_nb\_senior & 72.44 & -3.15 & $<.001$*** & -0.187 & Small \\
in\_w\_senior  & 73.12 & -2.47 & $<.001$*** & -0.164 & Small \\
in\_m\_senior  & 73.39 & -2.19 & $<.001$*** & -0.149 & Small \\
in\_nb\_senior & 73.45 & -2.14 & $<.001$*** & -0.143 & Negligible \\
sp\_nb\_senior & 73.25 & -2.34 & $<.001$*** & -0.135 & Negligible \\
\hline
\end{tabular}
\begin{tablenotes}
\scriptsize
\item Note. Mean Diff = difference from neutral baseline ($M = 75.59$). $\delta$ = Cliff's Delta. $p_{\text{corr}}$ = Bonferroni-corrected $p$-value. ***$p < .001$ after Bonferroni correction ($\alpha_{\text{adjusted}} = 0.00278$).
\end{tablenotes}
\end{table}


\subsubsection{Code Quality Metrics}
Pylint scores quantify adherence to Python style conventions and best practices on a 0-10 scale (higher is better). 
Across all conditions, Pylint scores exhibit substantially smaller variation than Maintainability Index. The neutral baseline achieves a moderate score ($M = 4.14$, $SD = 2.38$), which is exceeded by several persona conditions, though differences are modest in magnitude. \textit{Senior Spanish} personas achieve the highest scores (e.g., Spanish Man Senior: $M = 4.97$; Spanish Woman Senior: $M = 4.94$), corresponding to improvements of approximately 0.8 points over neutral. \textit{Indian} personas show the smallest deviations from baseline, with no statistically significant differences after correction. 

A Kruskal-Wallis test detects a statistically significant overall difference across persona conditions,
$H(18) = 76.04$, $p < .001$, with a small effect size ($\eta^2 = 0.004$). 
Bonferroni-corrected Mann-Whitney U tests identify 7 of 18 personas with significantly higher Pylint scores than the neutral baseline (Table~\ref{tab:pylint_comparisons}). All significant effects are positive, indicating fewer style violations under these personas. However, all associated effect sizes are negligible (Cliff’s $\delta \leq 0.13$), implying minimal practical differences despite statistical significance. 

Three weak patterns emerge. First, \textit{Spanish} personas account for the majority of significant improvements (5/7), followed by \textit{Brazilian} personas (2/7), while no Indian persona differs significantly from neutral. Second, \textit{Senior} personas are slightly overrepresented among significant results, though the seniority effect is far smaller than that observed for maintainability. Third, no consistent gender pattern emerges, with men, women, and non-binary personas appearing across both significant and non-significant groups.

\begin{table}[h]
\centering
\caption{Pairwise comparisons -- each persona vs. neutral baseline (Pylint score).}
\label{tab:pylint_comparisons}
\scriptsize
\begin{tabular}{lcccccc}
\hline
\textbf{Persona} & \textbf{Mean} & \textbf{Mean Diff} & \textbf{$U$} & \textbf{$p_{\text{corr}}$} & \textbf{$\delta$} & \textbf{Sig.} \\
\hline
\multicolumn{7}{l}{\textit{Significant Comparisons (Better than Neutral)}} \\
sp\_m\_senior   & 4.97 & +0.83 & 629,786 & $<.001$*** & +0.132 & Yes \\
sp\_w\_senior   & 4.94 & +0.80 & 624,617 & $<.001$*** & +0.125 & Yes \\
br\_w\_senior   & 4.78 & +0.65 & 616,731 & $<.001$*** & +0.109 & Yes \\
sp\_nb\_senior  & 4.71 & +0.57 & 612,218 & $<.001$*** & +0.102 & Yes \\
sp\_m\_junior   & 4.75 & +0.62 & 606,110 & .005**     & +0.090 & Yes \\
br\_m\_senior   & 4.70 & +0.56 & 603,913 & .009**     & +0.085 & Yes \\
sp\_nb\_junior  & 4.66 & +0.52 & 602,818 & .011*      & +0.084 & Yes \\
\hline
\multicolumn{7}{l}{\textit{Non-Significant Comparisons}} \\
br\_nb\_senior  & 4.46 & +0.32 & 592,890 & .133       & +0.065 & No \\
br\_m\_junior   & 4.58 & +0.44 & 591,786 & .157       & +0.064 & No \\
sp\_w\_junior   & 4.59 & +0.45 & 587,810 & .358       & +0.057 & No \\
br\_nb\_junior  & 4.46 & +0.32 & 586,434 & .407       & +0.056 & No \\
in\_m\_senior   & 4.41 & +0.27 & 583,204 & .890       & +0.048 & No \\
in\_w\_senior   & 4.37 & +0.23 & 580,052 & $>$.999    & +0.042 & No \\
in\_w\_junior   & 4.41 & +0.27 & 577,123 & $>$.999    & +0.038 & No \\
br\_w\_junior   & 4.43 & +0.30 & 574,730 & $>$.999    & +0.037 & No \\
in\_nb\_senior  & 4.27 & +0.13 & 575,326 & $>$.999    & +0.036 & No \\
in\_nb\_junior  & 4.30 & +0.17 & 573,762 & $>$.999    & +0.033 & No \\
in\_m\_junior   & 4.32 & +0.18 & 570,714 & $>$.999    & +0.026 & No \\
\hline
\end{tabular}
\begin{tablenotes}
\scriptsize
\item Note. Mean Diff = difference from neutral baseline ($M = 4.14$). $\delta$ = Cliff's Delta. All effect sizes negligible ($|\delta| < 0.147$). ***$p < .001$, **$p < .01$, *$p < .05$ after Bonferroni correction ($\alpha_{\text{adjusted}} = 0.00278$).
\end{tablenotes}
\end{table}


\subsubsection{Security Metrics}
Across the full dataset, Bandit flagged a total of 40 security issues only, corresponding to $<$0.2\% of all outputs.  Of these findings, all are classified as \textit{low} severity, with no \textit{medium} or \textit{high} severity vulnerabilities observed. The flagged issues are sparsely distributed across personas, with no persona condition exhibiting a concentration of security findings sufficient to support meaningful statistical comparison


\subsection{RQ3: Observed Effects of Demographic Persona-Induced Prompts on Open-Weight Model}
\label{sec:results-rq3}
To assess whether demographic persona prompts yield similar effects across different LLMs, we applied the same RQ1 and RQ2 analysis pipelines to responses generated by GPT-OSS-120B (an open-weight model).

\subsubsection{Prevalence of Demographic Markers Across Models}
\begin{figure*}[t]
    \centering
    \includegraphics[height=0.65\textheight, keepaspectratio]{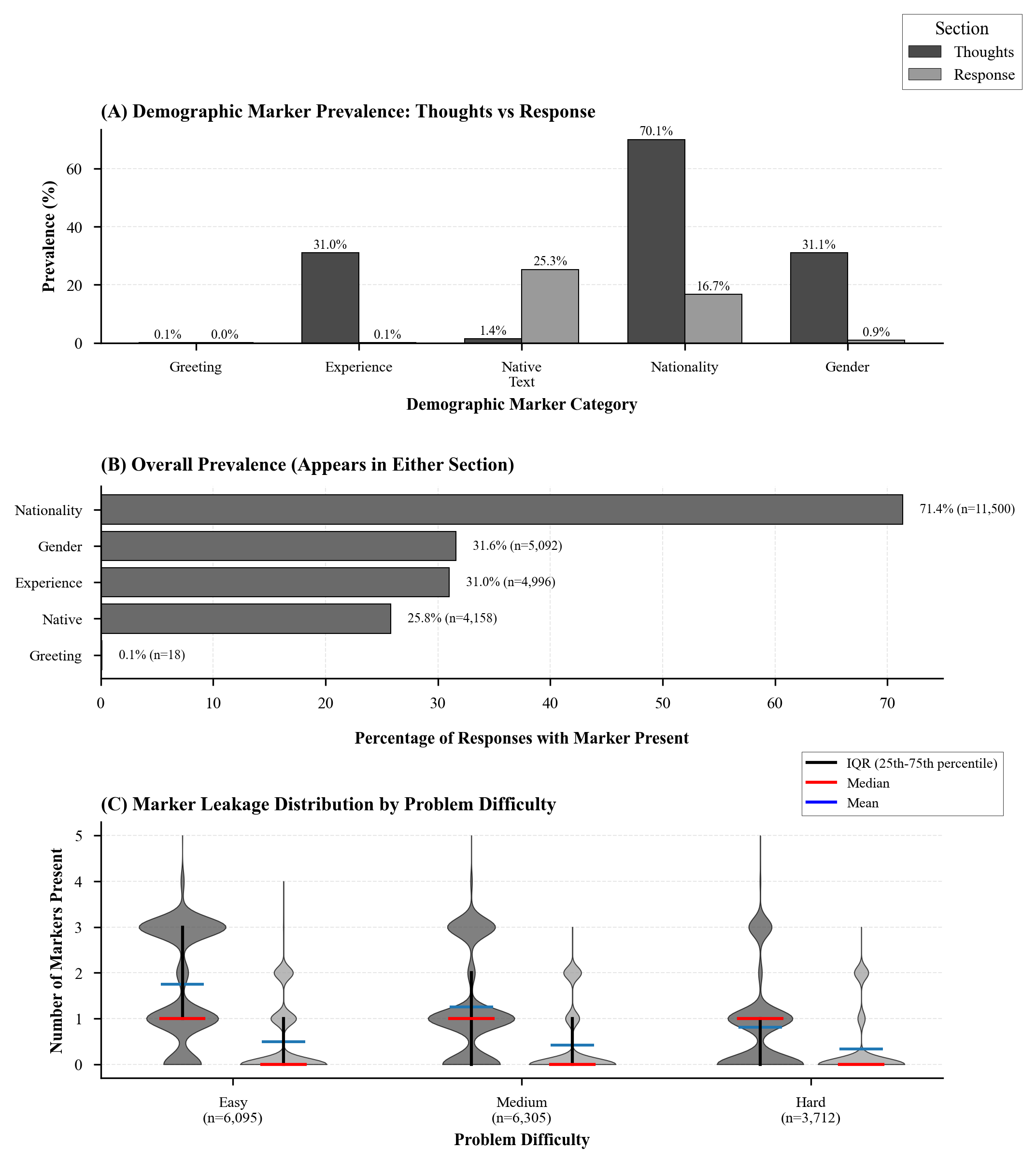}
    \caption{Demographic marker prevalence analysis for GPT-OSS-120B: 
    \textbf{(A)} Comparison of marker prevalence between \texttt{THOUGHTS} and \texttt{RESPONSE} sections, 
    \textbf{(B)} Overall prevalence appearing in either, and 
    \textbf{(C)} Distribution of marker counts stratified by problem difficulty.}
    \label{fig:RQ3Analysis}
\end{figure*}

We observed different marker prevalence patterns in GPT-OSS-120B compared to Gemini-2.5-pro, with differences in how markers manifest between \texttt{THOUGHTS} and \texttt{RESPONSE}, which can be seen within Figure \ref{fig:RQ3Analysis}. 

Figure \ref{fig:RQ3Analysis}-(A) showcases 
that \textit{NativeText} is the only marker that is amplified in GPT-OSS by increasing from 1.4\% in \texttt{THOUGHTS} to 25.3\% in \texttt{RESPONSE} (+23.9 percentage points). This pattern was also represented within Gemini-25-pro seen in Figure \ref{fig:RQ1Analysis1} which had an amplification from 9.9\% to 32.5\% (+22.6 percentage points). All other markers within GPT-OSS undergo suppression in \texttt{RESPONSE} relative to \texttt{THOUGHTS}, \textit{Nationality} declines from 70.1\% in \texttt{THOUGHTS} to 16.7\% (-53.4 percentage points), which is more than double that seen from Gemini-2.5 (-21.6 percentage points). \textit{Experience} drops from 31.0\% to 0.1\% (-30.9 percentage points) compared to Gemini's moderate decline of 41.1\% to 24.9\% (16.2 percentage points). \textit{Gender} falls from 31.1\% to 0.9\% (-30.2 percentage points) in GPT-OSS whereas Gemini showed minimal suppression (12.3\% to 9.5\%, -2.8 percentage points). 

Figure \ref{fig:RQ3Analysis}-(B) 
shows that \textit{Nationality} is the most prevalent marker, appearing in 71.4\% of responses relative to Gemini's 33.2\%. This ranking represents a fundamental shift from Gemini where \textit{Greeting} represented 65.6\% while GPT-OSS appeared in 0.1\% of responses. \textit{Gender} was referenced in 31.6\% of GPT-OSS responses, where Gemini had \textit{Gender} appearing in 19.2\%. \textit{Experience} occurred in 31.0\% of responses, while Gemini had 53.0\%. 
\textit{NativeText} appeared in 25.8\% of GPT-OSS's responses, where Gemini had 34.2\%. 

Figure \ref{fig:RQ3Analysis}-(C) shows the distribution of total marker counts across problem difficulty levels for GPT-OSS-120B. The median marker count remains at approximately 1.0 for \texttt{RESPONSE} section across \textit{easy}, \textit{medium}, and \textit{hard} problems. The IQR ranges show compression with increasing difficulty. The \texttt{THOUGHTS} section shows median counts of 1.5-2.0 markers across all difficulties. This relative stability contrasts with Gemini's pronounced difficulty effect, where marker distributions shifted substantially from \textit{easy} to \textit{hard} problems. Having said that, it is worth noting that 2936 \textit{hard} problems were excluded from the analysis as GPT{-}OSS could not generate a solution for them.

\subsubsection{Persona-Specific Leakage Patterns Across Architectures}

\begin{figure*}[t]
    \centering
    \includegraphics[height=0.50\textheight, keepaspectratio]{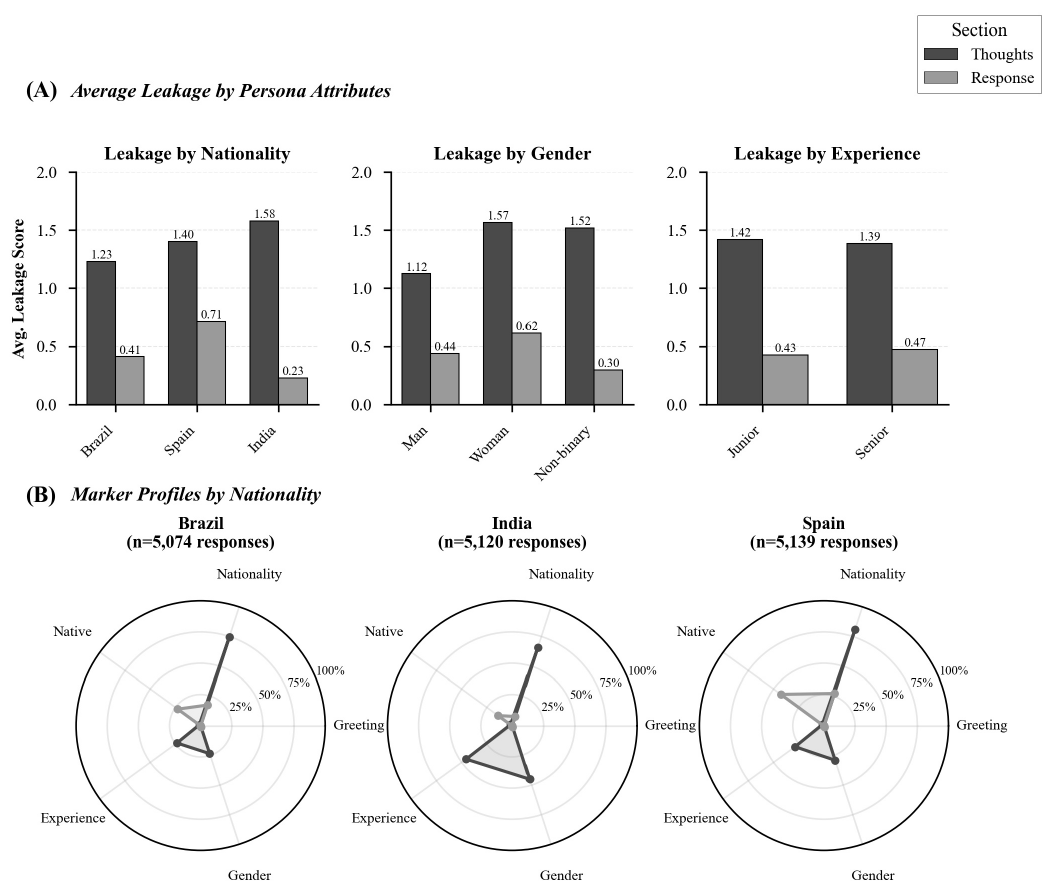}
    \caption{Persona-specific demographic marker leakage patterns for GPT-OSS-120B: \textbf{(A)} Aggregate leakage scores by persona attributes, \textbf{(B)} Spider charts by nationality.}
    \label{fig:RQ3Analysis2}
\end{figure*}

Figure~\ref{fig:RQ3Analysis2}-(A) shows 
that while Gemini amplifies nearly all persona dimensions from \texttt{THOUGHTS} to \texttt{RESPONSE}, GPT-OSS suppresses nearly all.

\textit{Nationality} rankings reverse between models: GPT-OSS ranks Spain highest (0.71), Brazil (0.41), India (0.23), while Gemini ranks Brazil (1.77), Spain (1.59), India (1.10). Both show India as lowest-leakage, but Gemini amplifies all nationalities (Brazil: +0.48, Spain: +0.37, India: +0.05), whereas GPT-OSS suppresses all (Brazil: -0.82, Spain: -0.69, India: -1.35). 
Gemini also amplifies both Junior (+0.27) and Senior (+0.34), with Junior showing higher final leakage (1.61 vs 1.37). GPT-OSS suppresses both (Junior: -0.99, Senior: -0.92). In \texttt{THOUGHTS}, Junior shows higher leakage (1.42 vs 1.39), but in \texttt{RESPONSE}, this reverses with Senior showing higher leakage (0.47 vs 0.43).

Figure~\ref{fig:RQ3Analysis2}-(B) reveals compositional inversion between models: GPT-OSS profiles are dominated by \textit{Nationality} (64-81\%), while Gemini profiles are dominated by \textit{Greeting} (~65-70\%). Spain shows highest \textit{Nationality} prevalence (~76\%) in GPT-OSS, while almost all nationalities showed no reference to \textit{Greeting}.


\subsubsection{Difficulty Effects Across Models}

\begin{figure*}[t]
    \centering
    \includegraphics[height=0.4\textheight, keepaspectratio]{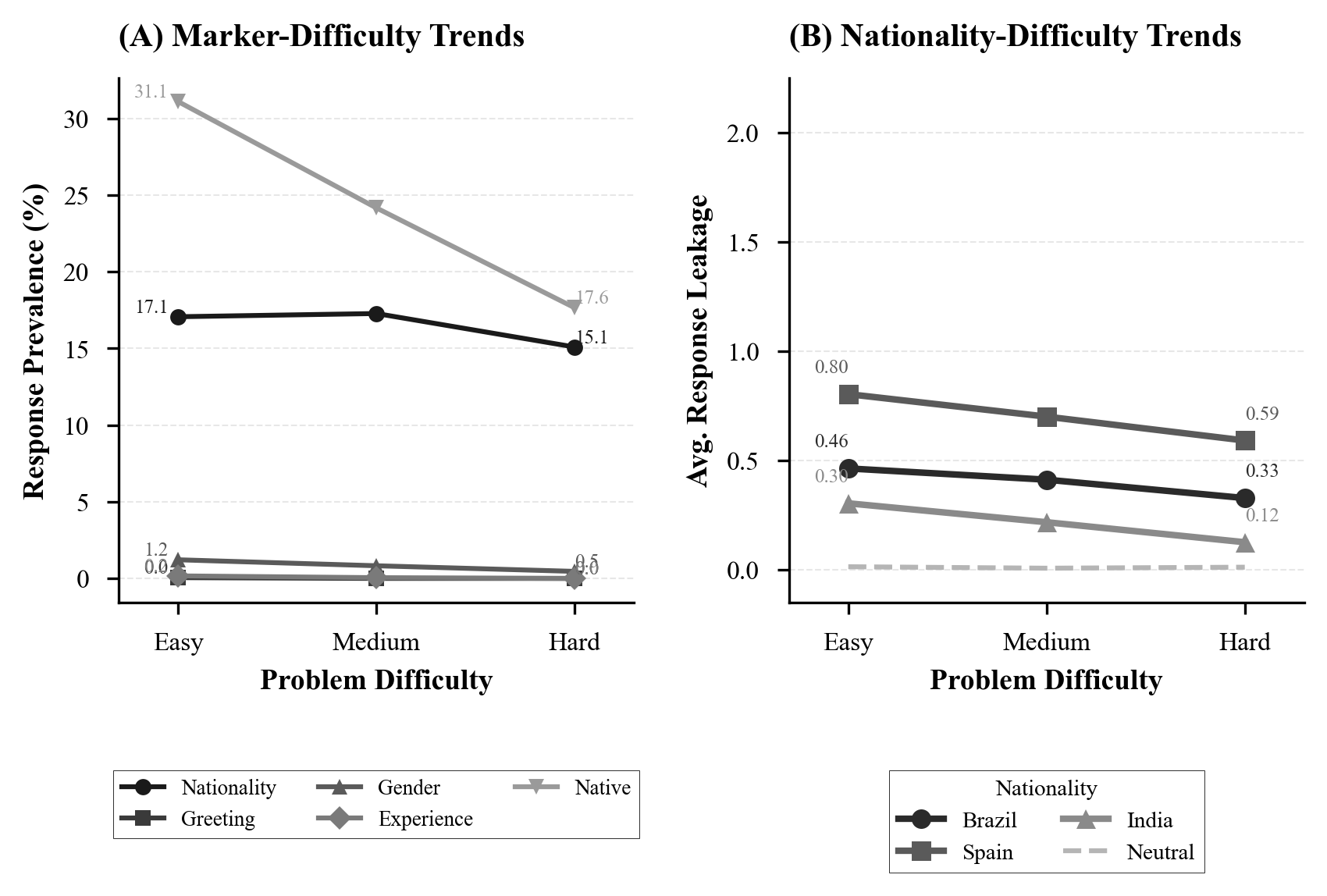}
    \caption{Difficulty effects on demographic marker leakage for GPT-OSS-120B: \textbf{(A)} Most markers show declining prevalence from easy to hard problems, and \textbf{(B)} Nationality-difficulty interaction shows a consistent decline across all 3 nationalities.}
    \label{fig:RQ3Analysis3}
\end{figure*}

Consistent with RQ1, where Gemini shows declines from \textit{easy} to \textit{hard} tasks in the demographic markers, 
Figure~\ref{fig:RQ3Analysis3} shows that all demographic markers in GPT-OSS-120B also decline as task difficulty increases. \textit{NativeText} shows the steepest decline from 31.1\% to 17.6\%. \textit{Nationality} remains relatively stable, declining only slightly from 17.1\% to 15.1\%, with slightly higher prevalence at \textit{medium} difficulty. All other markers remain at roughly 0\% across all difficulties. All three nationalities decline consistently in both models. Gemini shows Brazil declining from 2.02 to 1.61, Spain from 1.88 to 1.39, and India from 1.21 to 1.00. GPT-OSS shows Spain declining from 0.80 to 0.59, Brazil from 0.46 to 0.33, and India from 0.30 to 0.12. 

\subsubsection{Software Quality Evaluation Across Models}
To assess whether persona-induced effects on code quality generalize across architectures, we applied the RQ2 evaluation pipeline (Section \ref{sec:rq2-analysis}) to GPT-OSS-120B's generated solutions. 
We evaluated for correctness using LiveCodeBench's test harness, maintainability using the MI score, code style was evaluated with Pylint and security checks were compiled using Bandit.

\subsubsection{Functional Correctness}

We evaluated functional correctness across 1,055 LiveCodeBench problems for each of the 18 demographic persona conditions and the neutral baseline. The neutral baseline achieves an overall pass@1 accuracy of 36.68\%, which is significantly lower than Gemini's 67.58\%. 
For GPT-OSS-120B, all 18 personas exhibited higher accuracy than neutral, with improvements ranging from +3.41\% to +5.69\% (Table \ref{tab:gptoss_correctness}). The mean accuracy across all 18 personas for GPT-OSS was 41.20\%, which is an improvement of 4.52 percentage points relative to the baseline, while Gemini had an average decline of -1.54 percentage points. 

To assess statistical significance, we apply McNemar's test to each persona condition against the neutral baseline. The three correction methods (Bonferroni, Holm-Bonferroni, FDR) yielded identical conclusions: all 18 personas conditions show statistically significant improvements. Odds ratios range from 1.8 to 3.0, indicating that demographic persona prompting increases the likelihood of generating correct solutions by 80\% to 200\% relative to neutral prompting.

\begin{table}[htbp]
\centering
\caption{McNemar's test results (GPT-OSS-120B). A comparison of functional correctness of demographic personas against the neutral baseline.}
\label{tab:gptoss_correctness}
\begin{tabular}{lcccc}
\hline
\textbf{Persona} & \textbf{Persona Acc.} & \textbf{Diff (\%)} & \textbf{OR} & \textbf{p-value} \\
\hline
in\_nb\_senior    & 0.4237 & +5.69 & 3.000 & \textbf{0.000000} \\
sp\_w\_senior     & 0.4237 & +5.69 & 2.818 & \textbf{0.000000} \\
sp\_nb\_senior    & 0.4218 & +5.50 & 2.657 & \textbf{0.000000} \\
in\_w\_junior     & 0.4180 & +5.12 & 2.500 & \textbf{0.000002} \\
sp\_nb\_junior    & 0.4180 & +5.12 & 2.500 & \textbf{0.000002} \\
sp\_w\_junior     & 0.4180 & +5.12 & 2.459 & \textbf{0.000003} \\
sp\_m\_senior     & 0.4190 & +5.21 & 2.341 & \textbf{0.000004} \\
in\_w\_senior     & 0.4142 & +4.74 & 2.282 & \textbf{0.000015} \\
br\_m\_senior     & 0.4114 & +4.45 & 2.343 & \textbf{0.000021} \\
br\_w\_junior     & 0.4095 & +4.27 & 2.216 & \textbf{0.000055} \\
sp\_m\_junior     & 0.4095 & +4.27 & 2.154 & \textbf{0.000073} \\
in\_m\_senior     & 0.4085 & +4.17 & 2.100 & \textbf{0.000113} \\
br\_w\_senior     & 0.4066 & +3.98 & 2.077 & \textbf{0.000182} \\
br\_nb\_junior    & 0.4057 & +3.89 & 2.079 & \textbf{0.000217} \\
br\_nb\_senior    & 0.4085 & +4.17 & 1.957 & \textbf{0.000227} \\
br\_m\_junior     & 0.4019 & +3.51 & 2.088 & \textbf{0.000443} \\
in\_m\_junior     & 0.4038 & +3.70 & 1.867 & \textbf{0.000821} \\
in\_nb\_junior    & 0.4009 & +3.41 & 1.800 & \textbf{0.001821} \\
\hline
Mean              & 0.4120 & +4.52 & & \\
\hline
\end{tabular}
\begin{tablenotes}
\scriptsize
\item Neutral baseline accuracy = 0.3668. Bold \textit{p}-values indicate statistical significance after Bonferroni correction ($\alpha_{\text{adj}} = 0.00278$).
\end{tablenotes}
\end{table}

\subsubsection{Maintainability Metrics}
Maintainability was evaluated using the \path|radon| tool. Due to the exclusion of truncated responses from static analysis (Section~\ref{sec:rq3-analysis}), we analyzed 16,117 valid code samples across the 19 conditions. The neutral baseline achieved a mean MI score of 75.64 (SD = 9.63). In contrast to Gemini's findings, \textbf{no persona condition differs significantly from neutral} after Bonferroni correction (Table~\ref{tab:gptoss_maintainability}). 

A Kruskal-Wallis test indicates no significant overall difference across conditions for MI, $H(18) = 24.31$, $p = 0.145$, $\eta^2 < 0.01$. Pairwise Mann-Whitney U tests (Bonferroni-corrected, $\alpha = 0.00278$) show that all 18 demographic personas produce statistically insignificant MI scores from the neutral baseline (all $p_{corr} > 0.00278$; Table~\ref{tab:gptoss_maintainability}). Effect sizes (Cliff's $\delta$) are uniformly negligible, ranging from $\delta = 0.020$ to $\delta = 0.075$.

\begin{table}[htbp]
\centering
\caption{Pairwise comparisons: -- each persona vs. neutral baseline (maintainability index - GPT-OSS-120B).}
\label{tab:gptoss_maintainability}
\begin{tabular}{lccccl}
\hline
\textbf{Persona} & \textbf{Mean} & \textbf{Mean Diff} & \textbf{$p_{corr}$} & \textbf{$\delta$} & \textbf{Effect Size} \\
\hline
in\_m\_junior    & 76.85 & +1.21 & 0.187 & +0.075 & Negligible \\
br\_nb\_junior   & 76.67 & +1.03 & 0.283 & +0.071 & Negligible \\
br\_w\_junior    & 76.75 & +1.11 & 0.304 & +0.070 & Negligible \\
sp\_w\_senior    & 76.69 & +1.05 & 0.414 & +0.066 & Negligible \\
sp\_nb\_junior   & 76.49 & +0.86 & 0.437 & +0.066 & Negligible \\
br\_nb\_senior   & 76.58 & +0.94 & 0.654 & +0.061 & Negligible \\
br\_w\_senior    & 76.53 & +0.89 & 0.729 & +0.060 & Negligible \\
br\_m\_junior    & 76.50 & +0.86 & 0.759 & +0.060 & Negligible \\
sp\_w\_junior    & 76.64 & +1.00 & 0.767 & +0.059 & Negligible \\
sp\_m\_junior    & 76.54 & +0.90 & 0.871 & +0.058 & Negligible \\
sp\_nb\_senior   & 76.43 & +0.79 & 1.000 & +0.055 & Negligible \\
in\_w\_senior    & 76.37 & +0.74 & 1.000 & +0.048 & Negligible \\
in\_nb\_junior   & 76.23 & +0.59 & 1.000 & +0.044 & Negligible \\
in\_w\_junior    & 76.17 & +0.53 & 1.000 & +0.044 & Negligible \\
in\_m\_senior    & 76.27 & +0.63 & 1.000 & +0.039 & Negligible \\
br\_m\_senior    & 76.06 & +0.42 & 1.000 & +0.032 & Negligible \\
in\_nb\_senior   & 75.94 & +0.31 & 1.000 & +0.029 & Negligible \\
sp\_m\_senior    & 75.80 & +0.16 & 1.000 & +0.020 & Negligible \\
\hline
\end{tabular}
\begin{tablenotes}
\scriptsize
\item Note. Mean Diff = difference from neutral baseline (M = 75.64). $\delta$ = Cliff's Delta. $p_{corr}$ = Bonferroni-corrected p-value. No personas show significant differences after correction ($\alpha_{\text{adjusted}} = 0.00278$).
\end{tablenotes}
\end{table}

\subsubsection{Code Quality Metrics}
PyLint scores quantify adherence to Python style conventions and best practices on a 0-10 scale (higher is better). The neutral baseline achieves a mean score of 6.64 (SD = 1.28).

A Kruskal-Wallis test indicates no significant overall difference across persona conditions, $H(18) = 16.52$, $p = 0.555$, $\eta^2 < 0.01$. Bonferroni-corrected Mann-Whitney U tests reveal that \textbf{no persona differs significantly from neutral} (all $p_{corr} > 0.00278$; Table~\ref{tab:gptoss_pylint}). Effect sizes are uniformly negligible ($|\delta| < 0.15$), ranging from $\delta = -0.027$ to $\delta = +0.036$. In contrast, Gemini showed that 7 of 18 personas had statistically significant PyLint improvements (Table~\ref{tab:pylint_comparisons}). Both models show no practically meaningful stylistic differences, though Gemini exhibits detectable statistical variation where GPT-OSS-120B shows complete uniformity.

\begin{table}[htbp]
\centering
\caption{Pairwise comparisons -- each persona vs. neutral baseline (PyLint score - GPT-OSS-120B).}
\label{tab:gptoss_pylint}
\begin{tabular}{lcccccc}
\hline
\textbf{Persona} & \textbf{Mean} & \textbf{Mean Diff} & \textbf{U} & \textbf{$p_{corr}$} & \textbf{$\delta$} & \textbf{Sig.} \\
\hline
\multicolumn{7}{l}{\textit{Significant Comparisons (Better than Neutral)}} \\
\multicolumn{7}{l}{None} \\
\hline
\multicolumn{7}{l}{\textit{Non-Significant Comparisons}} \\
br\_w\_senior    & 6.65 & +0.01 & 310,171 & 1.000 & +0.036 & No \\
in\_m\_junior    & 6.67 & +0.03 & 311,716 & 1.000 & +0.031 & No \\
in\_w\_senior    & 6.68 & +0.04 & 318,894 & 1.000 & +0.028 & No \\
in\_nb\_senior   & 6.65 & +0.02 & 322,410 & 1.000 & +0.023 & No \\
br\_nb\_junior   & 6.66 & +0.02 & 313,179 & 1.000 & +0.022 & No \\
in\_nb\_junior   & 6.68 & +0.04 & 307,390 & 1.000 & +0.019 & No \\
br\_m\_junior    & 6.65 & +0.01 & 307,639 & 1.000 & +0.017 & No \\
br\_w\_junior    & 6.62 & $-0.02$ & 309,295 & 1.000 & +0.009 & No \\
br\_nb\_senior   & 6.64 & +0.00 & 303,438 & 1.000 & +0.008 & No \\
br\_m\_senior    & 6.60 & $-0.04$ & 310,064 & 1.000 & +0.008 & No \\
in\_w\_junior    & 6.64 & $-0.00$ & 310,892 & 1.000 & +0.003 & No \\
in\_m\_senior    & 6.60 & $-0.04$ & 305,536 & 1.000 & $-0.002$ & No \\
sp\_nb\_junior   & 6.60 & $-0.04$ & 309,192 & 1.000 & +0.000 & No \\
sp\_w\_junior    & 6.56 & $-0.08$ & 302,687 & 1.000 & $-0.014$ & No \\
sp\_nb\_senior   & 6.57 & $-0.07$ & 301,835 & 1.000 & $-0.019$ & No \\
sp\_m\_senior    & 6.56 & $-0.08$ & 305,022 & 1.000 & $-0.019$ & No \\
sp\_m\_junior    & 6.54 & $-0.10$ & 297,080 & 1.000 & $-0.020$ & No \\
sp\_w\_senior    & 6.51 & $-0.13$ & 299,770 & 1.000 & $-0.027$ & No \\
\hline
\end{tabular}
\begin{tablenotes}
\item Note. Mean Diff = difference from neutral baseline ($M = 6.64$). $\delta$ = Cliff's Delta. All effect sizes negligible ($|\delta| < 0.147$). ***$p < .001$, **$p < .01$, *$p < .05$ after Bonferroni correction ($\alpha_{\text{adjusted}} = 0.00278$).
\end{tablenotes}
\end{table}

\subsubsection{Security Metrics}
Across the 16,117 valid code responses output by GPT-OSS-120B, Bandit flagged a total of 916 security issues, corresponding to approximately 5.7\% of all outputs. The severity distribution shows 908 low-severity issues (99.1\%), 8 medium-severity issues (0.9\%), and 0 high-severity issues. Security findings are distributed across all persona conditions, ranging from 34 to 71 issues per persona, with the neutral baseline producing 48 issues.

GPT-OSS-120B responses exhibited higher security issue prevalence than Gemini-2.5. Both models show no systematic persona-specific patterns in security vulnerabilities, with findings distributed uniformly across conditions.

\section{Discussion and Implications}
\label{sec:discussion}

Our study analyzed whether demographic information of a user is associated with differences in LLM-generated reasoning and code. Our results show consistent associations between persona-induced prompts and (i) the presence of demographic markers in reasoning and responses, and (ii) variation in several measured code-quality dimensions.

Below, we discuss and interpret our findings and present our implications.

\textbf{Demographic Cues are Associated with Systematic Leakage in Reasoning and Responses (RQ1).}

Our findings for RQ1 show that demographic markers appear frequently in both reasoning traces and final responses, despite being unrelated to the programming problems themselves. This leakage was most pronounced for language-related markers (e.g., Native Text 33\% and Greetings 65\%), and substantially less common for explicit references to gender (19\%) or nationality (34\%) in final code outputs. Moreover, identity-linked markers (e.g., nationality, gender, experience) were more prevalent in \path|THOUGHTS|, while language-based markers were amplified in \path|RESPONSE|. 

This pattern suggests that while explicit identity cues tend to be attenuated in final outputs, potentially reflecting alignment or safety constraints, stylistic and linguistic adaptations remain more visible. Similar distinctions between reasoning and final outputs have been observed in prior work, where models use information internally that is partially filtered before presentation~\cite{wei2022chain,wu2025does}. In our context, this behaviour indicates that user demographic information is not ignored despite being irrelevant for the task, aligning with observations from persona-prompting studies that show increased internal variability even when outputs appear neutral~\cite{gupta2024biasrunsdeepimplicit,luz2025helpful}.

Moreover, leakage was more frequently observed for Brazilian and Spanish personas, junior developers, and binary gender identities, and declined monotonically with increasing task difficulty. One plausible interpretation is that LLMs treat demographic context as a signal for linguistic accommodation, adapting surface-level discourse features such as greetings, explanatory tone, or language choice. This behaviour aligns with prior work showing that LLMs often adjust their responses in light of additional contextual information to maximize perceived helpfulness or personalization \cite{luz-de-araujo-etal-2025-principled,magister2024wayllmpersonalizationlearning}. 

The decline in demographic leakage with increasing task difficulty suggests that persona-induced effects are not uniform across problem types. When tasks are harder, models appear to allocate more capacity to problem-solving and less to stylistic or contextual adaptation. This aligns with findings in prompt-engineering research showing that persona or role information has stronger effects in low-complexity tasks~\cite{luz-de-araujo-etal-2025-principled}. Indeed, this has a methodological implication for future bias evaluations: studies focused only on difficult benchmarks may underestimate persona effects that are more visible in simpler programming tasks.

\faHandORight~\underline{Implication 1:} \emph{User demographic information may influence internal reasoning even when final outputs appear superficially neutral.} This implies that safety mechanisms that aim to suppress references to sensitive attributes in user-facing responses may not fully eliminate demographic conditioning during problem solving.

\faHandORight~\underline{Implication 2:} \emph{LLMs appear to use demographic attributes of the user mainly to adjust language and style rather than technical content}. While such adjustments may be perceived as helpful or polite, they can introduce unnecessary variability and could lead to inconsistent user experiences.

\textbf{Associations Between Persona-Induced Prompts and Core Software Quality Metrics (RQ2).}
Across all evaluated dimensions, neutral prompts consistently outperform persona-induced prompts. All 18 personas are associated with lower functional correctness relative to a neutral baseline, with one persona (Indian, non-binary, junior) showing a statistically significant 3.6\% decrease after correction. The analysis of non-functional metrics reveals that persona-induced prompts were associated with reduced maintainability scores, with medium-to-large effect sizes concentrated among junior personas and Brazilian and Spanish personas. While some senior personas show small, statistically significant improvements in code style, these effects are negligible in magnitude. Security vulnerabilities are rare overall. The absence of clear security effects also echoes earlier findings that LLM security risks are more strongly driven by task framing and API usage than by stylistic or contextual prompts~\cite{dai2025rethinkingevaluationsecurecode}.

Among these results, the most robust and practically meaningful finding is the consistent reduction in maintainability under persona-induced prompts. Maintainability is a key determinant of long-term software evolution and technical debt, and its systematic decline in this setting may suggest that persona context may co-occur with code characteristics that are less prone to long-term modification and comprehension. One plausible explanation for this pattern is that persona-induced prompts or reasoning models encourage the generation of more verbose or pedagogically oriented code, particularly for junior personas. Such code may include additional branching or explanatory structure that preserves functional correctness and stylistic consistency, but increases algorithmic complexity, or control-flow depth that are directly penalized by maintainability metrics. 

This interpretation is further supported by the observed divergence between maintainability and PyLint scores. While PyLint primarily captures stylistic consistency and localized code issues, maintainability metrics reflect properties (algorithmic complexity, control-flow structure, and code volume) that directly affect long-term modifiability. 
The co-occurrence of stable or slightly improved PyLint scores with reduced maintainability suggests that user demographics may bias generation toward code that remains stylistically well-formed while becoming structurally more complex or verbose, thereby increasing maintenance burden without corresponding benefits to long-term modifiability.

While earlier persona-prompting studies emphasize performance gains from expert or role-based personas~\cite{kong-etal-2024-better,luz2025helpful}, our results align more closely with work showing that irrelevant personas introduce variability and sometimes degrade performance~\cite{luz-de-araujo-etal-2025-principled}. By embedding demographic attributes as contextual metadata rather than explicit role instructions, our study simulates real-world personalization features more closely than traditional role-play prompting.

\faHandORight~\underline{Implication 3:} \emph{User's demographic data may be associated with lower maintainability scores.} AI-assisted developer tools that store or infer user demographics risk unequal quality of assistance. If junior or certain national personas systematically receive less maintainable code, this could reinforce existing inequities in learning and productivity. Tool designers should consider: isolating technical generation from user demographic metadata, defaulting to neutral personas for core code generation, and auditing outputs for persona-induced variability. 

\textbf{Persona-Induced Effects Hold Across Model Families (RQ3).}
Both Gemini~2.5~Pro and GPT-OSS-120B exhibit demographic marker leakage in reasoning and responses, including nationality references, greetings, and experience-based framing. While absolute prevalence varies, there is evidence that demographic personas systematically affect the generated responses. In both cases, leakage is more prevalent in easier tasks, language-based markers are more likely to appear in final outputs, and neutral prompts minimize leakage.

This aligns with prior evidence that bias is not confined to a single model family or training paradigm \cite{gupta2024biasrunsdeepimplicit,liu2023uncovering,salewski2023incontextimpersonationrevealslarge}. Unlike prior cross-model comparisons that focus on explicit bias benchmarks or textual tasks, our results demonstrate that persona-induced bias in technical code generation exists across proprietary and open-weight models. This suggests a systemic issue rooted in training data and instruction-following behaviour rather than model-specific implementation details.

\faHandORight~\underline{Implication 4:} \emph{Understanding and mitigating persona-induced bias requires changes at the level of prompt design, tool architecture, and evaluation practices, not merely model selection.}

\textbf{Broader Implications for Software Engineering Practice.}

Our findings highlight an important research area in LLM-assisted software development: demographic information about users, even when included for benign personalization purposes, may subtly influence technical outputs. While the observed effects are generally small, they occur systematically and at scale, which raises concerns for long-term fairness and consistency in developer tooling.

For organizations deploying LLM-based coding assistants with memory or personalization features, our results suggest a need for careful consideration of what user attributes are stored, inferred, or surfaced to models during technical tasks. At the same time, our results do not suggest that demographic personalization is inherently harmful. Some forms of adaptation such as adjusting explanation depth based on experience may be beneficial when explicitly intended and transparently designed. The challenge, therefore, lies in distinguishing intentional, task-relevant personalization from unintended demographic data that add unnecessary variability.

\section{Conclusion}

We examined whether demographic information about users, when semantically irrelevant to the programming task, is associated with differences in LLM-generated code. Across three research questions, we observed consistent patterns in which persona-induced prompts are associated with LLM outputs, ranging from linguistic expressions to core technical properties of the generated code, even in purely algorithmic problem-solving settings. Demographic cues systematically influence LLM reasoning, impacting software maintainability and, in some cases, correctness, with effects observed across two model types we considered (Gemini Pro and GPT-OSS). By connecting work on bias, prompt design, and software quality, our work highlights a previously under-explored consideration for LLM-assisted software development. Our results point to the need for further research on robust and fair software engineering tools.

\section{Acknowledgements}

We would like to thank the Natural Sciences and Engineering Research Council of Canada (NSERC) Alliance and National Council for Scientific and Technological Development (CNPq) for funding this research. We would also like to thank Google for providing Gemini API research credits.

\section{Declarations}
\vspace{-1em}
The authors make the following declarations:

\vspace{-1.5em}

\subsection{Funding}
\vspace{-1em}
This research is partially supported by CNPq (420406/2023-9 and 442779/2023-2), and NSERC Alliance (ALLRP 592516 - 23)

\vspace{-1.5em}

\subsection{Ethical Approval}
\vspace{-1em}
Not applicable. This study did not involve human participants.

\vspace{-1.5em}

\subsection{Informed Consent}
\vspace{-1em}
Not applicable. No human data or personally identifiable information were collected or analyzed.

\vspace{-1.5em}

\subsection{Author Contributions}
\vspace{-1em}
\textbf{Anubhav Gupta} co-designed RQ1, RQ2 and RQ3 under the supervision of the Last Author; participated in the design, implementation, and validation of the prompt construction strategy; implemented RQ1 and RQ2 (formulated the scripts, results and analyses and involved with sensitivity analyses and qualitative coding); involved in manuscript writing and revision.\\
\\
\textbf{Mayara Costa Figueiredo} contributed to the conception and design of the study; participated in the design and validation of the prompt construction strategy; contributed to the design and implementation of RQ1 (sensitivity analyses and qualitative coding); contributed to manuscript writing and revision.\\
\\
\textbf{Let\'icia Santos Machado} contributed to the conception and design of the study; participated in the design and validation of the prompt construction strategy; contributed to the design and implementation of RQ1 (sensitivity analyses and qualitative coding); contributed to manuscript writing and revision. \\
\\
\textbf{Tanner Wright} contributed to the design and implementation of RQ3 under the supervision of the First and Last Author; involved in manuscript writing.\\
\\
\textbf{Ivan Beschastnikh} contributed to the formulation of the study’s initial idea and to manuscript writing and revision. \\
\\
\textbf{Cleidson R. B. de Souza} contributed to the formulation of the study’s initial idea and to manuscript writing and revision. \\
\\
\textbf{Gema Rodr\'iguez-P\'erez} contributed to the conception and design of the study; supervised RQ1, RQ2 and RQ3; participated in the design and validation of the prompt construction strategy; contributed to the design and implementation of RQ1 (sensitivity analyses and qualitative coding); contributed to manuscript writing and revision.\\
\\
All authors read and approved the final version of the manuscript.

\vspace{-1.5em}

\subsection{Data Availability Statement}
\vspace{-1em}
All scripts, datasets, and supplementary materials are available in our replication package~\cite{personatactics}.
\footnote{\url{https://osf.io/w2jfy/overview?view_only=fb0d5e26aef34f67852c9390cc9a7639}}

\vspace{-1.5em}

\subsection{Conflict of Interest}
\vspace{-1em}
The authors declare that they have no conflict of interest.

\vspace{-1.5em}

\subsection{Clinical Trial Number}
\vspace{-1em}
Not applicable.

\bibliographystyle{spmpsci}      
\bibliography{software}   

\clearpage
\section*{Authors and Affiliations}
Anubhav Gupta\orcidlink{0000-0003-0187-8192} \textbf{.}
Mayara Costa Figueiredo\orcidlink{0000-0001-8038-7409} \textbf{.}
Let\'icia Santos Machado\orcidlink{0000-0002-9105-3727} \textbf{.} 
Tanner Wright\orcidlink{0009-0003-2267-1622} \textbf{.}
Ivan Beschastnikh\orcidlink{0000-0003-1676-8834} \textbf{.}
Cleidson R. B. de Souza\orcidlink{0000-0003-3240-3122} \textbf{.}
Gema Rodr\'iguez-P\'erez\orcidlink{0000-0002-0062-8418}

\begingroup
\setlength{\parindent}{0pt}
\newcommand{\authblock}[3]{
  \textbf{#1} #2\\
  \texttt{#3}\par\vspace{0.9\baselineskip}
}

\vspace{1.2\baselineskip}

\authblock{\large Anubhav Gupta}%
{\\ University of British Columbia, Kelowna, BC, Canada}%
{anubhav.gupta@ubc.ca}

\authblock{\large Mayara Costa Figueiredo}%
{\\ Universidade Federal do Pará,  Belém, Brazil}%
{mcfigueiredo@ufpa.br}

\authblock{\large Let\'icia Santos Machado}%
{\\ Universidade Federal do Rio Grande do Sul, Porto Alegre, Brazil}%
{leticia.machado@inf.ufrgs.br}

\authblock{\large Tanner Wright}%
{\\ University of British Columbia, Kelowna, BC, Canada}%
{tanner.wright@ubc.ca}

\authblock{\large Ivan Beschastnikh}%
{\\ University of British Columbia, Vancouver, BC, Canada}%
{bestchai@cs.ubc.ca}

\authblock{\large Cleidson R. B. de Souza}%
{\\ Universidade Federal do Pará,  Belém, Brazil}%
{cleidson.desouza@acm.org}

\authblock{\large Gema Rodr\'iguez-P\'erez}%
{\\ University of British Columbia, Kelowna, BC, Canada}%
{gema.rodriguezperez@ubc.ca}
\endgroup

\end{document}